\documentclass[epsf,prb,twocolumn,showpacs,longbibliography,nofootinbib]{revtex4-1}
\usepackage{amsmath,amssymb,amsthm,amscd,latexsym,epsfig,bm}
\usepackage{float}
\usepackage[justification=raggedright,font=small]{caption}
\usepackage{graphicx}
\usepackage{color}
\usepackage{framed}
\usepackage[colorlinks,bookmarks=false,citecolor=blue,linkcolor=red,urlcolor=blue]{hyperref}
\usepackage{verbatim}
\usepackage[normalem]{ulem}

\newcommand{\zz}{\mathbb{Z}_2}
\newcommand{\z}{\mathbb{Z}}

\def\bA{\boldsymbol{A}}
\def\bE{\boldsymbol{E}}

\begin{document}

\title{Fracton phases via exotic higher-form symmetry-breaking}
\author{Marvin Qi}
\author{Leo Radzihovsky}
\author{Michael Hermele}
\affiliation{Department of Physics and Center for Theory of Quantum Matter, University of Colorado, Boulder, CO 80309, USA}
\date{\today}

\begin{abstract}
We study $p$-string condensation mechanisms for fracton phases from the viewpoint of higher-form symmetry, focusing on the examples of the X-cube model and the rank-two symmetric-tensor ${\rm U}(1)$ scalar charge theory.  This work is motivated by questions of the relationship between fracton phases and continuum quantum field theories, and also provides general principles to describe $p$-string condensation independent of specific lattice model constructions.  We give a perspective on higher-form symmetry in lattice models in terms of cellular homology.  Applying this perspective to the coupled-layer construction of the X-cube model, we identify a foliated 1-form symmetry that is broken in the X-cube phase, but preserved in the phase of decoupled toric code layers.  Similar considerations for the scalar charge theory lead to a framed 1-form symmetry.  These symmetries are distinct from standard 1-form symmetries that arise, for instance, in relativistic quantum field theory.  We also give a general discussion on interpreting $p$-string condensation, and related constructions involving gauging of symmetry, in terms of higher-form symmetry.
\end{abstract}

\maketitle

\tableofcontents

\section{Introduction}
\label{sec:intro}

Effective quantum field theories (QFTs) play a central role in the study of quantum phases of matter, as descriptions capturing the universal low-energy properties of a phase, and of critical points between phases.  While a great many quantum phases of matter are described at low energies by a known QFT, the general relationship between phases of quantum matter and QFTs is far from understood.  This has become particularly clear in recent years with the discovery of and increasing progress in understanding fracton phases of matter.\cite{chamon,haah11local,vijay15newkind,vijay16fracton,pretko17subdimensional,fractonarcmp,pretko20fracton}

By definition, fracton phases have gapped excitations for which no local operator can move an isolated excitation in one or more (or all) spatial directions, without creating additional excitations.  Such excitations are necessarily fractional, \emph{i.e.} single such excitations cannot be created locally.  Excitations confined to move along one-dimensional lines or two-dimensional planes are often referred to as lineons and planons, respectively, while fully immobile excitations are dubbed fractons.

Fracton phases present a challenge to our understanding of the relationship between phases of matter and QFTs for a number of reasons.  For instance, until recently it was a piece of conventional wisdom that any gapped phase of matter has a low-energy topological QFT (TQFT) description.  However, gapped fracton phases have a robust ground state degeneracy, where on a spatial $d$-torus (\emph{i.e.} with periodic boundary conditions) the dimension of the ground state subspace grows sub-extensively with system size.\cite{bravyi11topological, haah11local, haah13thesis}  (Here and throughout, $d$  denotes spatial dimension.)  This is incompatible with a TQFT description, which assigns a system-size-independent ground state subspace to the $d$-torus and other closed spatial manifolds.  Moreover, the number of superselection sectors (\emph{i.e.} distinct types of particle-like excitations) also diverges. \cite{haah13commuting, haah16algebraic, pai19fracton}  This is  suggestive of an infinite number of fields in the continuum.

Continuum quantum theories of fracton phases have been constructed and studied in a number of works \cite{pretko17subdimensional,slagle17XcubeQFT,slagle19symmetric,slagle19foliatedQFT,slagle20foliatedQFT,pretko18fracton-elasticityDuality,pretko18supersolid,pretko19crystal,gromov19chiral,gromov19multipole,gromov20duality,nguyen20fractonelasticity,bulmash18generalized,pretko18gaugeprinciple,you20ChernSimons,radzihovsky20vector,radicevic19systematic,wang19higherrank1,wang19higherrank2,wang19nonabelian,shenoy20fractonic,seiberg20vector, seiberg20exotic2d, seiberg20exotic3dU1, seiberg20exotic3dZN, gorantla20moreexotic}.  In general, these theories depend  strongly on ultra-violet scale physics, beyond the usual issues of regularization.  For instance, and more precisely, the theories of Refs.~\onlinecite{seiberg20exotic2d, seiberg20exotic3dU1, seiberg20exotic3dZN, gorantla20moreexotic} have a finite number of fields, but admit an uncountably infinite  number of discontinuous zero-action field configurations.  Such unusual properties of these theories do \emph{not} mean they are not useful; instead, this is an indication that much remains to be understood about the interplay between fracton lattice models on one hand, and continuum theories on the other.

Indeed, by showing that some fracton phases, dubbed foliated fracton phases, can be understood as renormalization group (RG) fixed points, the results of Ref.~\onlinecite{shirley18manifolds} suggest that continuum descriptions should be useful in capturing the universal properties of fracton phases.  The RG procedure employed in Ref.~\onlinecite{shirley18manifolds}  goes beyond the usual lore by allowing one to integrate out non-trivial lower-dimensional topological phases, further supporting the idea that continuum descriptions of fracton phases should lie in some sense beyond conventional QFT.

In this paper, we approach these issues from the starting point of fracton lattice models, with the aim of understanding universal properties of fracton phases from perspectives similar to those employed in QFT.  One path forward is provided by a number of existing concrete lattice-model constructions of fracton phases in terms of more conventional quantum phases of matter that do have a QFT description.  Some of these constructions begin with a collection of ordinary systems that are then coupled together, so that a fracton phase results upon condensation of some extended objects.\cite{ma17coupled,vijay17coupled,radzihovsky20vector}  Such mechanisms for fracton phases are referred to as $p$-string condensation, where ``$p$'' stands for particle, because the extended objects are strings built from point-like -- but not locally createable -- particle excitations.  In another construction of Williamson, Bi and Cheng, one starts with an ordinary gauge theory, and gauges some apparently ordinary global symmetries to obtain a fracton phase.\cite{williamson19SETfracton}  

The purpose of this paper is to study the above constructions, employing a perspective that intertwines higher-form symmetries and spatial geometrical structure.  Higher-form symmetries are a generalization of ordinary global symmetries whose role in QFT has attracted significant recent attention.\cite{gaiotto15generalized} We also note the earlier works of Refs.~\onlinecite{batista05Elitzur,nussinov09symmetry,nussinov09symmetry2,nussinov12holographies}, and more recently of Refs.~\onlinecite{weinstein19universality,weinstein20absence}, under the name of $d$-dimensional gauge-like symmetries.  Roughly speaking, a $q$-form symmetry ($q \geq 0$) has extended $q$-dimensional objects that carry the charge, while the symmetry operators are defined on $(d-q)$-dimensional spatial submanifolds.  $q=0$ is the case of an ordinary global symmetry.  $q$-form symmetry allows for systematic discussion of the condensation of extended objects in terms of spontaneous symmetry breaking.  This perspective arises naturally -- for $q=1$ -- in studying $p$-string condensation, which involves condensation of one-dimensional extended objects.

Our study of $p$-string condensation also turns out to be intimately tied to spatial geometry.  To understand what we mean by geometry, it is helpful to recall that in order to study a given QFT in arbitrary space-times, one usually needs to specify not only the space-time manifold ${\cal M}$, but also some geometrical structure on ${\cal M}$ such as a Riemannian metric, an orientation, a spin structure, and so on.  Many works have noted that geometry also plays a key role in fracton phases, and the geometrical structures arising tend to be more ``rigid'' than geometry arising in QFT.  For example, the  X-cube model has planon excitations that move only within $xy$, $yz$ and $xz$ planes.  This implies that the X-cube fracton phase cannot have a low-energy limit enjoying the full rotation symmetry of three-dimensional space, even allowing for fine-tuning.  In this case, the  geometrical structure can be precisely formulated as a certain kind of foliation structure on the \emph{spatial} manifold $M$.\cite{shirley18manifolds}  For many other fracton phases, including in particular the so-called rank-two symmetric tensor ${\rm U}(1)$ scalar charge theory (``scalar charge theory'' for short),\cite{pretko17subdimensional} a similar understanding is lacking.  Our treatment of $p$-string condensation in the scalar charge theory suggests that it may be possible to define this theory on spatial manifolds with a Riemannian metric and a choice of framing (see Sec.~\ref{sec:u1}).

An alternative motivation for this work is simply to understand the principles behind $p$-string condensation, going beyond a collection of specific lattice model constructions.  It is well-known that string condensation, where the strings are made out of locally createable excitations (\emph{i.e.}, made out of spins), is a mechanism for topological order.\cite{wen03artificial,wen03quantum,levin05stringnet}  This was first understood by studying simple models, but can also be understood in a model-independent fashion using higher-form symmetry.\cite{wen19highersymm}  Here, we develop a framework that allows us to give model-independent meaning to the condensation of $p$-strings built from non-locally-createable excitations.

One might wonder about the relevance of microscopic $q$-form symmetries, given that realistic models of solid state systems certainly do not have these symmetries without fine-tuning.  It is important to emphasize that the fracton phases resulting upon $p$-string condensation -- at least those studied in this paper -- remain robust upon explicitly breaking the $1$-form symmetry, or any symmetry for that matter.  Put another way, microscopic 1-form symmetries allow for a precise description of $p$-string condensation mechanisms.  The stability of the resulting phases to various perturbations is then a separate question to be addressed in its own right.  This situation is analogous to many familiar topologically ordered phases, \emph{e.g.} in the $\zz$ toric code model, which can be understood in terms of condensation of 1-dimensional objects charged under a microscopic 1-form symmetry (see Sec.~\ref{subsec:hfs-toric-code}), but which are robust to arbitrary perturbations without imposing any symmetry conditions.

We now outline the remainder of the paper, which begins in Sec.~\ref{sec:hfs} with a discussion of higher-form symmetry.  While this discussion is mostly a review, some ideas that are new (to our knowledge) are presented, and even readers familiar with higher-form symmetries are encouraged to consult Sec.~\ref{sec:hfs} to familiarize themselves with our perspective and notational conventions.  A more detailed outline of the subsections within Sec.~\ref{sec:hfs} is provided at the beginning of the section.

Section~\ref{sec:xcube} discusses the $p$-string condensation route to X-cube fracton order from the point of view of breaking a foliated 1-form symmetry.  In Sec.~\ref{sec:xcube-pstring}, we discuss $p$-string condensation in the coupled-layer construction of the X-cube model, generalized to an arbitrary multifoliated spatial manifold.  The foliated 1-form symmetry is introduced in this context, and analyzed in more detail in the following two sections.  First, Sec.~\ref{sec:charged} discusses charged operators that create $p$-strings, while Sec.~\ref{sec:breaking} studies  breaking of the foliated 1-form symmetry.  The symmetry-breaking is precisely characterized by studying ground states of the X-cube model on a multifoliation of the spatial 3-torus that corresponds to periodic boundary conditions twisted by the lattice translation symmetry.  Next, Sec.~\ref{subsec:variants} discusses variants of the construction of Sec.~\ref{sec:xcube-pstring}, which involve gauging a certain subgroup of a faithful $\z_2$ 1-form symmetry (see Sec.~\ref{subsec:hfs-general} for a definition of the term faithful in this context).  Finally, in Sec.~\ref{sec:XCIR} we discuss the relationship between the foliated 1-form symmetry and an emergent symmetry arising in the infrared, dubbed the ${\cal S}$-symmetry, generated by appropriate string logical operators.  While the ${\cal S}$-symmetry is not part of the foliated 1-form symmetry, we argue that any local Hamiltonian invariant under the foliated 1-form symmetry necessarily also enjoys the ${\cal S}$-symmetry.

Following the discussion of the X-cube model, in Sec.~\ref{sec:generalized}, we step back and give a general, abstract description of $p$-string condensation in terms of 1-form symmetry breaking.  We then apply this discussion to the ${\rm U}(1)$ scalar charge theory in Sec.~\ref{sec:u1}.  First, the construction of the scalar charge theory in terms of gauging symmetry is described from the higher-form symmetry point of view (Sec.~\ref{sec:u1-gauging}). Here, the notion of framed 1-form symmetry is introduced, and it is suggested that it may be possible to define the scalar charge theory on framed spatial manifolds with a Riemannian metric.  Unlike the case of foliated 1-form symmetry, we do not yet have a precise characterization of the breaking of framed 1-form symmetry, and we make some remarks about this issue in Sec.~\ref{sec:u1-gauging}.  This is followed by a brief discussion of the $p$-string construction of the scalar charge theory in Sec.~\ref{sec:u1-pstring}.  Finally, the paper concludes with a discussion of open issues in Sec.~\ref{sec:discussion}.

\section{Higher-form symmetry}
\label{sec:hfs}

In this section, we discuss those aspects of higher-form symmetries\cite{gaiotto15generalized} that pertain to our discussion of $p$-string condensation and related gauging constructions.  (Again, see also Refs.~\onlinecite{batista05Elitzur,nussinov09symmetry,nussinov09symmetry2,nussinov12holographies,weinstein19universality,weinstein20absence}.)  Much of the literature on higher-form symmetry is focused on relativistic QFT, while our discussion focuses on many-body lattice models in $d$ spatial dimensions, and therefore we encourage readers already familiar with higher-form symmetry at least to skim this section, to familiarize themselves with our notation and terminology.  Higher-form symmetries in lattice models have been discussed previously in a number of works.\cite{yoshida16generalized, wen19highersymm, tsui20lattice}

While much the discussion in this section is a review of known results, the distinction between faithful and non-faithful higher-form symmetries (Sec.~\ref{subsec:hfs-general}), and the related cellular homology viewpoint on higher-form symmetries (Sec.~\ref{subsec:cellular}), are new to our knowledge, and will play an important role.   We  begin with a general discussion of higher-form symmetry in Sec.~\ref{subsec:hfs-general}.  We then proceed in Sec.~\ref{subsec:hfs-toric-code} to illustrate this  discussion with the concrete example of the $d=2$ toric code, then in Sec.~\ref{subsec:cellular} providing a viewpoint on higher-form symmetry in terms of cellular homology.  Section~\ref{subsec:hfs-charged} discusses charged objects and the breaking of higher-form symmetry.

\subsection{General discussion}
\label{subsec:hfs-general}

Let $G$ be a group.  An ordinary unitary internal $G$ global symmetry, also referred to as a 0-form symmetry, acts on Hilbert space via unitary operators supported on all of space.  That is, given an element $g \in G$, there is a unitary operator $U(g)$ supported on all of space, and these unitaries form a  (possibly projective) representation of $G$.  We always assume that $U(g)$ is local in a sense that we do not try to define precisely, but rather illustrate by examples:  The simplest possibility is that $U(g)$ is a product of  unitary operators, each of which acts on a single lattice site.  A more general possibility is for $U(g)$ to be a finite-depth quantum circuit; this arises when describing the action of symmetry on the spatial boundary of symmetry protected topological phases.\cite{chen11edge,levin12braiding,else14classifying}

We now move to higher-form symmetries, which are always Abelian. Therefore, instead of a general group $G$, we consider an Abelian group denoted $A$.   An $A$ $q$-form symmetry has symmetry operators acting on arbitrary closed, non-empty submanifolds of space $M$ with co-dimension $q$, denoted $M^{d-q}$.  That is, given a closed submanifold $M^{d-q} \subset M$ and an element $a \in A$, there is a unitary operator $U(a; M^{d-q})$ supported on $M^{d-q}$.
Moreover, fixing $M^{d-q}$, these operators form a linear representation of $A$:
\begin{equation}
U(a; M^{d-q}) U(a'; M^{d-q}) = U(a + a'; M^{d-q}) \text{.} \label{eqn:rep}
\end{equation}
It is also assumed that all the unitaries defined above commute, \emph{i.e.}
\begin{equation}
U(a; M^{d-q}) U(a'; N^{d-q}) =  U(a'; N^{d-q}) U(a; M^{d-q}) \text{.}
\end{equation}
The unitaries $U(a; M^{d-q})$ generate the symmetry; more general symmetry transformations are products of the generating unitaries, and in general can have support on non-manifold subspaces of $M$.  As for 0-form symmetry, we assume that the $U(a, M^{d-q})$ symmetry operators are local; for instance, they could be products of single-site unitaries over submanifolds $M^{d-q}$.  Unless $A$ is a direct sum of $\zz$'s, then in general $-a \neq a$ for $a \in A$, and the submanifolds $M^{d-q}$ need to be specified together with an orientation.  We have $U(-a, M^{d-q}) = U(a, \bar{M}^{d-q} )$, where $\bar{M}^{d-q}$ is $M^{d-q}$ with the opposite orientation.

Here and throughout the paper, we assume that the spatial manifold $M$ is orientable.  There may be interesting global phenomena having to do with the orientability (or lack thereof) of $M$; exploring such issues is left for future work.

The objects carrying charge under a 0-form symmetry are point-like, \emph{i.e.} they are local operators that insert $G$-charge.  Roughly speaking, for a $q$-form symmetry the charged objects are extended $q$-dimensional objects defined on $q$-dimensional submanifolds of space.  We  discuss charged objects in more detail in Sec.~\ref{subsec:hfs-charged}.

It is important to distinguish between microscopic higher-form symmetries that are exact at all energy scales, and higher-form symmetries that act within some low-energy subspace of the Hilbert space.  We refer to the former symmetries as UV higher-form symmetries, and the latter as IR higher-form symmetries.  In a system with a UV higher-form symmetry, we may obtain a non-trivial IR higher-form symmetry by projecting the UV symmetry into a low-energy subspace, for instance the ground state subspace.  Even with no UV higher-form symmetry, IR higher-form symmetries can emerge when there is an energy gap to breaking the extended charged objects -- an emergent IR higher-form symmetry holds at energy scales below such a gap.  When discussing IR higher-form symmetries, the operator $U(a; M^{d-q})$ should be understood as acting within an appropriate low-energy subspace.

We now introduce the notion of a \emph{faithful} $q$-form symmetry, which will play an important role in our subsequent discussions.  Informally, for a faithful $q$-form symmetry, different group elements in $A$ and/or different submanifolds $M^{d-q}$ give rise to distinct symmetry operators.  More formally, we first observe that $U(0; M^{d-q}) = 1$ for all $M^{d-q}$, by Eq.~(\ref{eqn:rep}).  Now let $a, a' \in A$ with at least one of $a$ or $a'$ non-zero, and let $M^{d-q}, N^{d-q} \subset M$ be two non-empty closed $(d-q)$-submanifolds.  We say the $q$-form symmetry is faithful if $U(a; M^{d-q}) = U(a', N^{d-q})$ implies that $a = a'$ and $M^{d-q} = N^{d-q}$.  Fixing $M^{d-q}$, this implies that $U(a, M^{d-q})$ gives a faithful representation of $A$, which justifies the terminology (see also below in Sec.~\ref{subsec:hfs-toric-code}).

Some intuition for the notion of a faithful $q$-form symmetry is provided by the example of a ${\rm U}(1)$ gauge theory with and without dynamical matter degrees of freedom, and with electric field denoted by $E$.  We consider the electric 1-form symmetry, with symmetry operators
\begin{equation}
U(\theta; M_{d-1}) = \exp\Big( i \theta \Phi_E(M_{d-1}) \Big) \text{,} \label{eqn:electric-1form}
\end{equation}
where $\theta \in [0, 2\pi)$ and $\Phi_E(M_{d-1})$ is the operator measuring the electric flux through $M_{d-1}$.  If we consider the pure gauge theory with Gauss' law $\nabla \cdot E = 0$, then $U(\theta; M_{d-1}) = 1$ whenever $M_{d-1}$ is a boundary of some $d$-dimensional region.  Evidently in this case we have a non-faithful 1-form symmetry.  Moreover, we can say the 1-form symmetry is topological, in the sense that $U(\theta; M_{d-1})$ does not change if we make smooth deformations of $M_{d-1}$.  More formally, we can say that $U(\theta; M_{d-1}) = 1$ whenever the homology class (with $\z$ coefficients) of $M_{d-1}$ is trivial, and $U(\theta; M_{d-1})$ only depends on the homology class of $M_{d-1}$.

Now we can instead consider a theory with dynamical charged matter with density $\rho$, so that Gauss' law becomes $\nabla \cdot E = \rho$.  Such a theory can still be invariant under the electric 1-form symmetry, with symmetry operators still given by Eq.~(\ref{eqn:electric-1form}).  Imposing this symmetry makes the charged matter completely immobile in space.  Now when $M_{d-1}$ bounds some $d$-dimensional region $R$, \emph{i.e.}  $M_{d-1} = \partial R$, then
$\Phi_E(M_{d-1}) = \int_R \rho$, the charge enclosed in $R$.  Therefore $U(\theta; M_{d-1})$ is never the identity operator as long as $\theta \neq 0$, and the 1-form symmetry is faithful.

Ref.~\onlinecite{seiberg20vector} comments on the relationship between this terminology and terminology more standard in the quantum field theory literature on higher-form symmetries.  In a relativistic quantum field theory, $q$-form symmetries are topological and are thus never faithful.  However in a non-relativistic setting, faithful symmetries are a natural possibility.

It is useful to make some further remarks about where faithful and non-faithful $q$-form symmetries tend to arise.  Faithful $q$-form symmetries tend to arise as UV symmetries of lattice models with a tensor product Hilbert space (although it will be very important for us that not all such $q$-form symmetries are faithful).  For instance,  in the example of ${\rm U}(1)$ gauge theory with dynamical matter, if one considers a lattice regularization then in fact one has a system with a tensor product Hilbert space -- the electric field can vary arbitrarily on different lattice links, and the Gauss' law constraint can be viewed as simply defining a new field $\rho$ that tracks some aspects of the electric field configuration.  In the same context of lattice models with tensor product Hilbert space, topological (non-faithful) $q$-form symmetries arise more naturally as IR symmetries.  This would occur for instance in a system with emergent ${\rm U}(1)$ gauge structure and gapped electric charge excitations, at energy scales below the charge gap.

\subsection{Toric code example}
\label{subsec:hfs-toric-code}

We give a concrete example of a 1-form symmetry in the $d=2$ $\zz$ toric code model on the square lattice.\cite{kitaev2003fault}  1-form symmetries of this model have been discussed previously.\cite{wen19highersymm} We consider a $L \times L$ square lattice with periodic boundary conditions (\emph{i.e.} we take the spatial manifold $M = T^2$, the 2-torus) in the limit $L \to \infty$.   The vertices are labeled by $v$ and nearest-neighbor links by $\ell$.  On each link we place a qubit, denoting $x$ and $z$ Pauli operators by $X_\ell$ and $Z_\ell$.  The Hamiltonian is
\begin{equation}
H_{TC} = - \sum_v A_v - \sum_p B_p - h \sum_\ell Z_\ell \text{,}  \label{eqn:htc}
\end{equation}
where $p$ labels square plaquettes, $B_p = \prod_{\ell \in p} X_\ell$ with the product taken over the links in the perimeter of $p$, and
$A_v = \prod_{\ell \ni v} Z_\ell$, with the product taken over the four links $\ell$ touching the vertex $v$.  When $h=0$ the Hamiltonian is exactly solvable as a sum of commuting terms.  For $h < h_c$, the model is in a gapped phase with topological order whose universal properties are captured by the $h=0$ solvable point.

It will be useful to recall that the toric code can be mapped exactly to a $\zz$ lattice gauge theory coupled to $\zz$-charged matter.  The $\zz$ gauge theory has qubits on links with Pauli operators $\tilde{Z}_\ell$ and $\tilde{X}_\ell$, and matter fields on sites with $x$ and $z$ Pauli operators $\tau^x_v$ and $\tau^z_v$.  There is a Gauss' law constraint
\begin{equation}
\prod_{\ell \ni v} \tilde{Z}_\ell = \tau^z_v \text{.}
\end{equation}
The mapping between the toric code model and the $\zz$ gauge theory can be understood in terms of the following operator dictionary:
\begin{eqnarray}
Z_\ell &=& \tilde{Z}_\ell \\
X_\ell &=& \tau^x_{v_1} \tilde{X}_\ell \tau^x_{v_2}  \text{,}
\end{eqnarray}
where $v_1$ and $v_2$ are the vertices on the ends of $\ell$.  The $\zz$ gauge theory Hamiltonian is thus
\begin{equation}
H_{{\rm gauge}} = - \sum_v \tau^z_v - \sum_p B_p - h \sum_\ell \tilde{Z}_\ell \text{,}
\end{equation}
where now $B_p = \prod_{\ell \in p} \tilde{X}_\ell$.  We see that $\tilde{Z}_\ell$ is the electric field, while $\tilde{X}_\ell$ is the vector potential.  Moreover, $\tau^z_v$ measures the $\zz$ charge at $v$, while $\tau^x_v$ creates/destroys $\zz$ charge at $v$ (and is thus not a gauge-invariant operator).  While this Hamiltonian has dynamical matter, it is immobile, \emph{i.e.} $\tau^z_v$ is a constant of the motion.\footnote{Here the term dynamical refers to the fact that the matter degrees of freedom are described by a dynamical field, as opposed to a background field with no dynamics.}  This would be changed by adding a term like
$- t \sum_\ell \tau^x_{v_1} \tilde{X}_\ell \tau^x_{v_2}$, which is a kinetic energy for the $\zz$ matter, but we shall not add this term.  (In the toric code language, this term is $-t \sum_\ell X_\ell$, \emph{i.e.} it is a transverse Zeeman magnetic field.)  In gauge theory language, $h < h_c$ ($h > h_c$) is the deconfined (confined) phase. With $h=0$ we have a fine-tuned point in the deconfined phase with zero bare tension for electric field lines.

\begin{figure}
        \includegraphics[width=0.7\columnwidth]{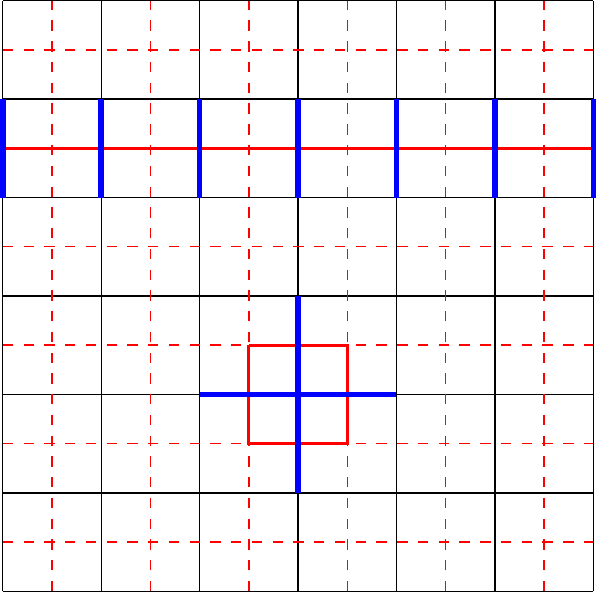}
        \caption{Illustration of the direct (black) and dual (dashed red) lattice for the toric code. Two choices of $M^1$ are shown as solid red lines in the dual lattice; the corresponding operators $U(M^1)$ are supported on the blue links $\bar{\ell}$ intersecting $M^1$. The $U(M^1)$ for the top operator corresponds to a cycle with nontrivial homology class, and the $U(M^1)$ for the bottom operator is $A_v$. }
        \label{fig:tclattice}
\end{figure}

Returning to the toric code model, we discuss the electric 1-form symmetry, where the terminology comes from the mapping to $\zz$ gauge theory.  However, because the model enjoys an electric-magnetic self-duality, we find this terminology somewhat confusing, and we will refer to this symmetry as the $Z$ 1-form symmetry, or simply the $Z$-symmetry when its 1-form character is clear from context.  The symmetry operators are defined on closed curves in the dual lattice (see Fig.~\ref{fig:tclattice}), whose links $\bar{\ell}$ are in one-to-one correspondence with the links $\ell$ of the original lattice that they cross, so we can write \emph{e.g.}
$Z_{\bar{\ell}} = Z_\ell$.  The symmetry operator on the manifold $M^1$ is
\begin{equation}
U(M^1) = \prod_{\bar{\ell} \in M^1} Z_{\bar{\ell}} \text{,}  \label{eqn:tc-symmetry}
\end{equation}
where $M^1$ is taken to be a set of dual links $\bar{\ell}$ forming a closed curve, and we omit writing the unique non-trivial element of $\zz = \{0 ,1\}$, \emph{i.e.}  $U(M^1) \equiv U(1; M^1)$.  Under the mapping to the gauge theory, $U(M^1)$ measures the $\zz$ electric flux through $M^1$.  These operators are easily seen to commute with the Hamiltonian, and moreover $A_v = U(M^1(v))$, where $M^1(v)$ is the smallest dual closed curve encircling the vertex $v$.  Adding a kinetic energy term for the $\zz$ matter breaks the symmetry; this is why we did not add such a term.

We note that if we set $h = 0$, the model also has a magnetic $\zz$ 1-form symmetry (or $X$-symmetry), where the symmetry generators are products of $X_\ell$ on closed curves in the original (not dual) lattice.  The electric and magnetic 1-form symmetries are exchanged by electric-magnetic duality.

Returning to the $Z$-symmetry, this is a UV symmetry and is easily seen to be faithful; if $M^1 \neq N^1$, then $U(M^1) \neq U(N^1)$ because these are two different products of $Z_\ell$'s.  At the $h=0$ point, we can easily project the UV 1-form symmetry to the ground state subspace, obtaining a non-faithful -- and topological -- IR 1-form symmetry.  This happens because $A_v = 1$ within the ground state subspace, so that $U(M^1)$ only depends on the homology class of $M^1$ in $H_1(M, \zz)$.  For instance $U(M^1)$ is non-trivial if $M^1$ wraps around one of the cycles of the spatial 2-torus.  In fact, the same IR 1-form symmetry holds not just in the ground state subspace, but below the gap to $A_v = -1$ excitations (we can make this gap the largest scale in the problem by making the coefficient of $A_v$ in the Hamiltonian large).  In gauge theory language, this subspace corresponds to the pure gauge theory, with Gauss' law constraint $\prod_{\ell \ni v} \tilde{Z}_\ell = 1$.  If we start with a pure gauge theory (rather than obtaining it as a low-energy effective model), then this non-faithful 1-form symmetry is a UV symmetry.

\subsection{Cellular homology viewpoint on $q$-form symmetries}
\label{subsec:cellular}

We now present a viewpoint on $q$-form symmetries in terms of cellular homology, which is useful when studying lattice models. (We refer the reader to Ref.~\onlinecite{hatcherbook} for an introduction to cellular homology.)  Another motivation for this viewpoint is the following:  in our discussion so far, we emphasized the role of $q$-form symmetry operators  supported on $d-q$-manifolds.  However, this emphasis is a bit unnatural, as the product of two symmetry operators $U(a, M^{d-q}) U(a', N^{d-q})$ is supported on the space $M^{d-q} \cup N^{d-q}$, which can have non-manifold points where $M^{d-q}$ and $N^{d-q}$ intersect.  Instead, we will describe $q$-form symmetry operators as supported on (cellular) cycles in homology, which allows for self-intersecting spaces.  We first give a general discussion of this viewpoint, then illustrate it in the context of the $d=2$ toric code example described above.

We consider a $d$-dimensional spatial manifold $M$, which we give a cell structure.  We consider cellular homology with coefficients in an Abelian group $A$, and let $C_n(M, A)$ be the group of cellular $n$-chains with $A$ coefficients.  We have the usual boundary operators $\partial_n : C_n(M, A) \to C_{n-1}(M, A)$ satisfying $\partial_{n-1} \partial_n = 0$, and $\partial_0 = 0$ by convention.   As usual we drop the subscript on $\partial_n$ when its value is clear from the context.  The groups of cellular boundaries and cycles are $B_n(M, A) = \operatorname{Im} \partial_{n+1}$ and $Z_n(M, A) = \operatorname{Ker} \partial_n$, with the $n$th homology group given by the quotient $H_n(M, A) = Z_n(M, A) / B_n(M, A)$.

We use the cell structure to introduce a quantum lattice model, for instance by attaching qubits to some of the cells, and taking a tensor product to obtain the Hilbert space ${\cal H}$. If desired, after taking the tensor product, we can also impose a local constraint such as a Gauss' law in a gauge theory. We denote the group of unitary operators acting on ${\cal H}$ by ${\cal U}({\cal H})$.  

To describe a $q$-form symmetry with symmetry group $A$, we identify the $q$-form symmetry group ${\cal A}^q \equiv Z_{d-q}(M, A)$, the group of cellular $(d-q)$-cycles.  We then define a $q$-form symmetry to be a unitary representation
\begin{equation}
U : {\cal A}^q \to {\cal U}({\cal H}) \text{,}
\end{equation}
that is, we assume $U$ is a group homomorphism.  In addition we assume this representation is local in the following sense.  If $c_{d-q} \in {\cal A}^q$, we can talk about the support of $c_{d-q}$ in space $M$.  For instance, a simple case is where $c_{d-q}$ models a $(d-q)$-dimensional submanifold of $M$.  Without being too picky about precise definitions, we assume that  support of the unitary $U(c_{d-q})$ agrees with that of the cycle $c_{d-q}$.  

In this language it is simple to describe the faithful and topological $q$-form symmetries that arose in the discussion above.  A faithful $q$-form symmetry is one where $U$ is a faithful representation (\emph{i.e.} $U$ is an injective group homomorphism), and distinct cycles are represented by distinct symmetry operators.  On the other hand, a topological $q$-form symmetry is one where $U(c_{d-q}) = 1$ if and only if $c_{d-q} = \partial c_{d-q+1}$, \emph{i.e.} if $c_{d-q}$ is the boundary of a higher-dimensional region. For a topological $q$-form symmetry, only the homology class of the cycle $c_{d-q}$ matters in determining the corresponding symmetry operator $U(c_{d-q})$.  We observe that by these definitions, as expected, a topological $q$-form symmetry cannot arise as a UV symmetry in a product Hilbert space, because it must be the case that operators supported on different regions (on two different cycles representing the same homology class) have the same action on Hilbert space.

An advantage of this viewpoint on $q$-form symmetry is that it allows us to concretely understand various constructions in usual group-theoretic terms.  For instance, subgroups of a $q$-form symmetry are just subgroups of ${\cal A}^q$, and upon specifying a subgroup we can take the quotient by it.  We will need to consider examples where \emph{e.g.} a 0-form symmetry arises as a subgroup of a 1-form symmetry, which is straightforward to describe precisely in this language.

We remark that usually when discussing symmetry of a physical system, if a symmetry group has elements that act trivially on the system, the usual practice is to take the quotient by such elements to correctly identify the symmetry group.  In the case of a topological $q$-form symmetry, this results in a symmetry group $H_{d-q}(M, A) = Z_{d-q}(M,A) / B_{d-q}(M,A)$.  While this practice is useful, we do not follow it in this paper, because it sacrifices information about the spatial support of symmetry operators.

Now we illustrate this general discussion with the specific example of the $Z$-symmetry of the $d=2$ toric code model.  We take $M = T^2$, and take the obvious cell structure on $T^2$ given by the dual square lattice.  In particular, the 0-cells are the dual vertices, 1-cells are the dual links $\bar{\ell}$, and 2-cells are dual plaquettes.  The 1-form symmetry group is ${\cal Z}^1_2 \equiv Z_1(T^2, \zz)$.   A general element $c_1 \in Z_1(M, \zz)$ can be written
\begin{equation}
c_1 = \sum_{\bar{\ell}} n_{\bar{\ell}} \bar{\ell} \text{,}
\end{equation}
where the sum is over 1-cells $\bar{\ell}$, and the coefficients $n_{\bar{\ell}} \in \zz = \{0,1\}$ are arbitrary subject to the condition $\partial c_1 = 0$.  It is clear that $c_1$ can always be written as a sum of non-intersecting closed curves in the dual lattice. The faithful 1-form symmetry is given by the unitary representation
\begin{equation}
U(c_1) = \prod_{\bar{\ell}} (Z_{\bar{\ell}})^{n_{\bar{\ell}}} \text{.}
\end{equation}
If we impose the Gauss' law constraint $A_v = 1$ for each vertex $v$, then this faithful 1-form symmetry becomes a topological 1-form symmetry.

When $A = {\rm U}(1)$, we find it more convenient to consider the Hermitian operators generating the ${\rm U}(1)$ symmetry, which can be achieved by working in homology with $\z$ coefficients.  A ${\rm U}(1)$ 1-form symmetry is then a representation
\begin{equation}
\Phi : Z_{d-q}(M, \z) \to \operatorname{Herm}({\cal H}) \text{,}
\end{equation}
where $\operatorname{Herm}({\cal H})$ is the additive group of Hermitian operators acting on ${\cal H}$.

To illustrate this more concretely, consider a ${\rm U}(1)$ lattice gauge theory with a 1-form electric symmetry on a $d$-dimensional spatial lattice, with $(d-1)$-cells $\ell^{d-1}_i$, and with operators $E(\ell^{d-1}_i)$ with integer eigenvalues, measuring the electric flux through $\ell^{d-1}_i$.   A $(d-1)$-cycle $c_{d-1} \in Z_{d-1}(M, \z)$ can be written $c_{d-1} = \sum_i n_i \ell^{d-1}_i$, and is represented by the Hermitian operator $\Phi(c_{d-1}) = \sum_i n_i E(\ell^{d-1}_i)$, which measures the electric flux through $c_{d-1}$. The unitary symmetry operators are just exponentials $\exp(i \theta \Phi(c_{d-1}) )$.

\subsection{Charged operators and breaking of higher-form symmetry}
\label{subsec:hfs-charged}

For the purposes of this paper, the main use of $q$-form symmetry (with $q=1$) will be to identify and study the condensation of $1$-dimensional extended objects.  To that end, in this subsection we discuss the objects carrying charge under a $q$-form symmetry, and then discuss breaking of higher-form symmetry, which is associated with condensation of the charged objects.  We focus primarily on faithful UV $q$-form symmetries, but make some comments on other cases.

Roughly speaking, we can say the charge is carried by $q$-dimensional extended objects, but some care is needed to understand the precise meaning of that statement.  First, for a 0-form symmetry, let ${\cal O}_{\lambda}$ be an operator with support on some bounded region of space.  Taking $G$ to be Abelian to facilitate comparison with the $q > 0$ case, ${\cal O}_\lambda$ carries a definite charge under the 0-form symmetry if
\begin{equation}
U(g) {\cal O}_\lambda = \lambda(g) {\cal O}_\lambda U(g) \text{,} \label{eqn:0form-transformation}
\end{equation}
where $\lambda(g) \in {\rm U}(1)$ forms a one-dimensional representation of $G$.  We can summarize this situation by saying that the charged objects of a 0-form symmetry are point-like.

Now for $q > 0$, we consider an operator ${\cal O}_\lambda$ supported on a $q$-dimensional submanifold $M^q \subset M$, possibly with boundary.  Again, $\lambda(a)$ is a one-dimensional representation of $A$.  $M^q$ needs to be taken oriented unless $A$ is a direct sum of $\zz$'s.  We assume that ${\cal O}_\lambda$ is local in the same sense as the symmetry operators; for instance, it could be a product of single-site operators over $M^q$.  Considering a symmetry operator defined on $M^{d-q}$, we assume that $M^q$ and $M^{d-q}$ intersect transversally, so that the intersections are isolated points.  The analog of Eq.~(\ref{eqn:0form-transformation}) is
\begin{equation}
U(a; M^{d-q}) {\cal O}_\lambda = \Lambda {\cal O}_\lambda U(a; M^{d-q}) \text{,} \label{eqn:qform-transformation}
\end{equation}
where, by locality, the phase factor $\Lambda \in {\rm U}(1)$ is determined by the intersections between $M^{d-q}$ and $M^q$.
We assume that
\begin{equation}
\Lambda = \prod_i [\lambda(a)]^{o(i)} \text{,} \label{eqn:lambda-form}
\end{equation}
with the product over intersections between $M^{d-q}$ and $M^q$, and where the quantity $o(i) = \pm 1$ compares the orientation of $M^{d-q}$ and $M^q$ locally at the intersection $i$.  We say that ${\cal O}_\lambda$ carries definite charge under a $q$-form symmetry if Eqs.~(\ref{eqn:qform-transformation}) and ~(\ref{eqn:lambda-form}) hold for all $M^{d-q}$ and all $a$.

In the above example of the $d=2$ $\zz$ toric code model, the symmetry operator Eq.~(\ref{eqn:tc-symmetry}) acts on Wilson lines
\begin{equation}
O_\lambda = \prod_{\ell \in M^1} X_{\ell} \text{,}
\end{equation}
with $\lambda$ the unique non-trivial one-dimensional representation of $\zz$, and $M^1$ taken to be a set of links $\ell$ forming a closed curve. The relation Eq.~(\ref{eqn:qform-transformation}) can be verified directly since $X_\ell$ and $Z_{\bar{\ell}} = Z_\ell$ anticommute.

For a faithful UV $q$-form symmetry, we expect that given a representation $\lambda(a)$ and a $q$-submanifold $M^q$, a charged operator ${\cal O}_{\lambda}$ supported on $M^q$ can always be constructed.  To see this, consider a symmetry operator defined on a small $(d-q)$-sphere $S^{d-q}$, which can be thought of as enclosing a $(q-1)$-dimensional object.  This symmetry operator will act non-trivially on ${\cal O}_{\lambda}$ if $S^{d-q}$ encloses some part of the $(q-1)$-dimensional boundary $\partial M^q$.  For a faithful symmetry, any two different spheres $S^{d-q}$ give different symmetry operators, and it should be possible to find $M^q$ and ${\cal O}_{\lambda}$ on which the action of symmetry differs.  This will be the case if we can choose $M^q$ to be an arbitrary $q$-disc; then we can use symmetry operators defined on different spheres $S^{d-q}$ to detect the position of the boundary $\partial M^q$.  Once charged objects can be constructed on arbitrary $q$-discs, then it should be possible to define them on arbitrary $q$-submanifolds by gluing discs together.

If we consider instead a topological UV $q$-form symmetry, which is not faithful, then this situation changes.  A good example to have in mind is the electric 1-form symmetry of a ${\rm U}(1)$ gauge theory without matter, where electric field lines cannot end, and there is no gauge-invariant operator defined on a finite line segment that inserts an electric field line along the segment. More generally, we observe that $U(a; S^{d-q}) = 1$ for any small sphere $S^{d-q}$.  This immediately implies charged operators can only be constructed on $M^q$ without boundary.  For if we had a charged operator on some $M^q$ with non-trivial boundary,  choosing $S^{d-q}$ to enclose a part of the boundary would give a non-trivial transformation of ${\cal O}_{\lambda}$ under $U(a; S^{d-q})$, which is a contradiction.  Instead, in this case, we expect that charged operators can be defined on arbitrary $M^q$ without boundary.  Such considerations will play a role in understanding the charged objects of the other non-faithful symmetries that we encounter in this paper.

Returning to the case of a faithful symmetry, we thus see that there are variety of different charged objects, with different geometries and topologies.  If $M^q$ is a closed $q$-manifold, and especially if $M^q$ is large in size, it is natural to view ${\cal O}_\lambda$ as a $q$-dimensional extended object.  However if $M^q$ is small in size and has a boundary, it can be better to view ${\cal O}_\lambda$ as a point-like object.  We note that if $M^q$ is closed and is the boundary of some $(q+1)$-dimensional submanifold of space, then the charged operator ${\cal O}_\lambda$ has $\Lambda = 1$ for all symmetry operators $U(a; M^{d-q})$.  Such ${\cal O}_{\lambda}$, which we refer to as \emph{contractible} charged operators, do not transform under the symmetry at all.

Contractible charged operators are allowed to appear as terms in a local Hamiltonian invariant under the $q$-form symmetry, and these terms can drive condensation of extended $q$-dimensional charged objects and corresponding spontaneous breaking of the $q$-form symmetry.  When these terms appear in the Hamiltonian, there are dynamical processes where $q$-dimensional charged objects are created and destroyed, and where these objects fluctuate.  If the magnitude of such terms is sufficiently large, the charged objects may proliferate and condense.  It should be noted that the $q$-form symmetry forbids charged operators defined on $M^q$ with boundary from appearing as terms in the Hamiltonian, so we really have a condensation of $q$-dimensional extended objects, even in situations (such as a faithful $q$-form symmetry) where more general charged objects exist.

As with 0-form symmetries, condensation of the charged objects leads to spontaneous breaking of the symmetry.  One diagnostic for spontaneous breaking of a $q$-form symmetry is to consider the projection of symmetry and charged operators into the ground state subspace.  If this projection is non-trivial, the symmetry is broken.

The $d=2$ $\zz$ toric code phase is characterized by the spontaneous symmetry breaking of its $1$-form symmetry. Let us consider the fully deconfined limit $h=0$ on a spatial torus. The $4$-dimensional ground state subspace can be decomposed into the eigenbasis of the charged operators $\tilde{X_1} = O^1_{\lambda}$ and $\tilde{X_2} = O^2_{\lambda}$ supported on two inequivalent noncontractible cycles. The symmetry operators $\tilde{Z_1} = U(M^1)$ and $\tilde{Z_2} = U(M^{1'})$, with $M^1$ and $M^{1'}$ also inequivalent non-trivial cycles in homology, act
 nontrivially on this subspace, satisfying the Pauli algebra $\tilde{X_i}\tilde{Z_j} = (-1)^{\delta_{ij}} \tilde{Z_j} \tilde{X_i}$. Just as in the $0$-form symmetry case, this means that the degenerate ground states (in the $\tilde{X}_i$ eigenbasis) are labeled by the expectation values of the charged operators, and acting with a symmetry operator on a ground state shifts these expectation values.

On the other hand, if the projection is trivial, this does \emph{not} necessarily mean the symmetry is unbroken.  We can see this by considering the $d=2$ toric code model on a 2-sphere, where the ground state is unique, and the symmetry and charged operators have trivial projection to the one-dimensional ground state subspace.  This is evidently a subtle diagnostic for symmetry breaking because, unlike in the case of 0-form symmetry, it is tied to the global spatial topology.

A different diagnostic for symmetry breaking comes from correlation functions of charged operators. For a $0$-form symmetry, the two-point correlation functions decay differently in the symmetry broken and symmetry preserving phase. In the symmetry preserving phase, correlation functions decay exponentially to zero with a characteristic correlation length. In contrast, the two-point correlation function approaches a constant at long distance in the symmetry-broken phase.

For general $q$-form symmetries, analogous behavior can be seen in the ground state expectation values of ${\cal O}_\lambda$ for $M^q$ large and contractible with linear size $s$. In the symmetry preserving phase, $\langle {\cal O}_\lambda \rangle$ decays exponentially as $e^{- c s^{q+1} }$, while in the symmetry broken phase, the decay still exponential but slower, going like $e^{- c' s^q}$. In the context of the electric 1-form symmetry of pure gauge theories, this is the familiar  ``area-law" vs ``perimeter-law" behavior, which distinguishes confined and deconfined phases, respectively.

\section{X-cube fracton phase via foliated 1-form symmetry breaking}
\label{sec:xcube}

\subsection{$p$-string condensation and foliated 1-form symmetry}
\label{sec:xcube-pstring}

In this section we describe the $p$-string condensation route to the $\zz$ X-cube fracton phase,\cite{ma17coupled, vijay17coupled} from the point of view of breaking a 1-form symmetry.  The original discussions of $p$-string condensation involved the construction of a model on the simple cubic lattice, resulting in the X-cube model defined on the same lattice in a strong coupling limit.  Here, we take a QFT-inspired perspective where we start with a spatial manifold $M$, which is then endowed with some geometrical structure required for our constructions to make sense.  Informed by Ref.~\onlinecite{shirley18manifolds}, we will consider $M$ to be equipped with a certain foliation structure (described below), and develop $p$-string condensation in this setting from the point of view of breaking 1-form symmetry.

The relevant 1-form symmetry turns out to be different from the faithful and topological cases discussed in Sec.~\ref{sec:hfs}.  Instead, information about the foliation structure is encoded in the 1-form symmetry, which we thus refer to as a foliated 1-form symmetry.  The cellular homology point of view on $q$-form symmetry (Sec.~\ref{subsec:cellular}) plays a crucial role in delineating the distinctions among different types of 1-form symmetry.

We consider a closed spatial 3-manifold $M$.  The starting point for the $p$-string condensation route to the X-cube model is a system consisting of three stacks of $d=2$ toric code layers.  Right away it is clear that there is no natural way to place stacks of $d=2$ layers on $M$ without some geometrical structure.  Following Ref.~\onlinecite{shirley18manifolds}, we endow $M$ with a structure that we refer to as a \emph{multifoliation}, loosely following Ref.~\onlinecite{lawson1974foliations} (see p. 406).  We emphasize that we are describing the same geometrical structure that arises in Ref.~\onlinecite{shirley18manifolds}, but we use slightly different terminology that we feel is clearer and more descriptive.  In the mathematics literature, a foliation of a closed manifold $M$ is usually understood to mean a decomposition of $M$ as a disjoint union of closed submanifolds of lower dimension, with certain smoothness properties assumed.  The submanifolds of a foliation are called leaves, and the foliation is non-singular when all the leaves are of the same dimension.  For instance, we can visualize three-dimensional Euclidean space decomposed into the set of all $xy$ planes.  In a singular foliation, some isolated leaves are submanifolds of different dimension, or are non-manifold subspaces.  In the present case, we are interested in foliations of the 3-manifold $M$ by two-dimensional leaves.  Moreover, we always assume the leaves to be compact.  While our discussion extends to singular foliations as considered in Ref.~\onlinecite{shirley18manifolds}, for simplicity, we work only with non-singular foliations.

Following Ref.~\onlinecite{shirley18manifolds}, we need not just one foliation of $M$, but a set of three different foliations. Moreover, we assume that the leaves have ``nice'' intersection properties: (1) Any two 2-dimensional leaves intersect transversally or not at all; \emph{i.e.} their intersections is a one-dimensional submanifold where one leaf ``cuts through'' the other.  Note that two leaves can only intersect if they belong to two different foliations of $M$.  (2) Any three 2-dimensional leaves intersect at discrete points or not at all.  We refer to all of this structure as a multifoliation of $M$, or in more detail as a 3-multifoliation of $M$ by two-dimensional leaves, where the ``3'' indicates the number of different foliations of $M$ considered.  This contrasts with the terminology in Ref.~\onlinecite{shirley18manifolds}, which referred to the presence of the same structure as a  singular compact total foliation of $M$.  We prefer the term multifoliation, because this makes it very clear that more than one foliation of $M$ is involved, something that is  essential for the constructions that follow.

\begin{figure}
	\includegraphics[width=\columnwidth]{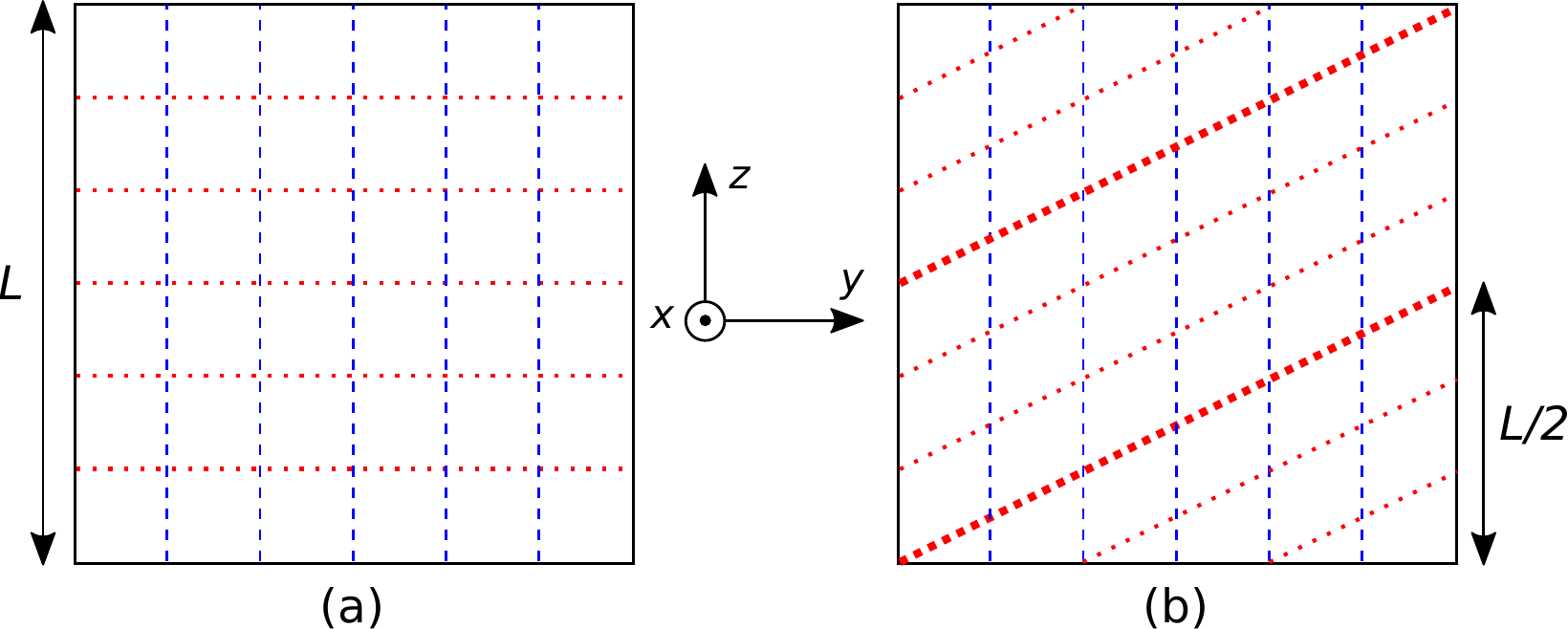}
	\caption{(a) Standard foliation structure and a discretization on $T^3$, shown for $L = 6$.  $T^3$ is viewed as a $L \times L \times L$ cube with opposite faces identified, and a $yz$-plane cross section of the cube is shown.  The dotted lines (red online) are cross sections of $xy$-plane leaves belonging to one foliation, while the dashed lines (blue online) are cross sections of $xz$-plane leaves belonging to another.  The $yz$-plane leaves are parallel to the cross section and are not shown.  (b) The twisted foliation structure on $T^3$ discussed in Sec.~\ref{sec:breaking}.  The $xz$ and $yz$-plane leaves are as in (a), but the $xy$-plane leaves are replaced with tilted $\widetilde{xy}$ leaves as indicated by dotted lines (red online).  Again $L = 6$, but due to the periodic boundary conditions there are only $L/2 = 3$ $\widetilde{xy}$-leaves in the discretization; the thick dotted line indicates a cross-section of a single $\widetilde{xy}$ leaf, which is homeomorphic to $T^2$. \label{fig:foliations}}
\end{figure}

Ref.~\onlinecite{shirley18manifolds} considered several examples of  multifoliated manifolds.  To illustrate our general discussion, it will often be enough to keep in mind the simplest concrete example, which is illustrated in Fig.~\ref{fig:foliations}(a).  Considering $M = T^3$, viewed as a cube in ${\mathbb R}^3$ with opposite faces identified, we take the three foliations to be the sets of all $xy$, $yz$ and $xz$ planes contained in $T^3$.  These leaves are thus squares with opposite sides identified, \emph{i.e.} each leaf is homeomorphic to $T^2$.  We refer to this as the standard multifoliation structure on $T^3$.  A different multifoliation structure on $M = T^3$, referred to as the twisted multifoliation structure, is shown in Fig.~\ref{fig:foliations}(b) and is discussed and used in Sec.~\ref{sec:breaking}.

Next, we choose a discretization of the multifoliation structure, which means that we choose a (non-empty) finite subset of two-dimensional leaves for each of the three foliations.  This defines a cell structure on $M$, where each 2-cell belongs to a single leaf, 1-cells lie in the intersection of two leaves, and 0-cells are the points where three leaves intersect.  3-cells are the ``voids'' filling the space between the leaves.  For the standard multifoliation structure on $T^3$, we think of $T^3$ as a $L \times L \times L$ cube with opposite faces identified, where $L$ is a positive integer.  Then we choose $L$ leaves in each foliation, equally spaced from their neighbors in the normal direction.  The resulting cell structure is nothing but a $L \times L \times L$ simple cubic lattice with periodic boundary conditions. Indeed, in general, the local environment of a 0-cell is the same as that of a vertex in the cubic lattice.

The X-cube model was originally defined on the simple cubic lattice,\cite{vijay16fracton} and later generalized to a discretized multifoliation structure,\cite{shirley18manifolds} as we now describe.  We place a qubit on each 1-cell $\ell$, with $x$ and $z$ Pauli operators denoted $Z_\ell$ and $X_\ell$.  The foliated X-cube Hamiltonian is
\begin{equation}
H_{XC} = - \sum_{v, \mu} A^{\mu}_v - \sum_{c_3} B_{c_3} \text{.}  \label{eqn:foliated-Xcube}
\end{equation}
The sum on $v$ is over 0-cells (vertices), and $\mu$ runs over the three foliations.  The operator $A^{\mu}_v$ is a product of the four $Z_\ell$ Pauli operators touching $v$ and contained within the $\mu$ foliation leaf that contains $v$.  The sum on $c_3$ is over 3-cells, with $B_{c_3} = \prod_{\ell \in c_3} X_\ell$, \emph{i.e.} the product is over the edges contained in the boundary of $c_3$.  For the standard foliation structure on $T^3$, the 3-cells are the elementary cubes of the simple cubic lattice, and $B_{c_3}$ is a product of $X_\ell$ over the 12 edges in the boundary of the cube $c_3$.  Because $[A^{\mu}_v, B_{c_3}] = 0$, this Hamiltonian is a sum of commuting Pauli operators and is exactly solvable.  It is in a gapped fracton phase with robust ground state degeneracy that grows with the system size (number of leaves in the discretized foliation), and excitations of restricted mobility; these properties have been studied on the simple cubic lattice\cite{vijay16fracton} and on more general foliated manifolds.\cite{shirley18manifolds}

Because we are interested in the $p$-string condensation route to the X-cube phase, rather than studying the foliated X-cube Hamiltonian of Eq.~(\ref{eqn:foliated-Xcube}), we instead consider the generalization of the coupled-layer construction of Refs.~\onlinecite{ma17coupled, vijay17coupled} to a general foliation.  Restricting the cell structure to a single leaf also gives a cell structure on the leaf, which can be used to place the $d=2$ toric code Hamiltonian Eq.~(\ref{eqn:htc}) on each leaf, taking $h = 0$.  In Sec.~\ref{subsec:hfs-toric-code} we discussed the same model on the square lattice, and very little needs to be changed from the discussion there. Here, qubits are placed on each 1-cell $\ell$, $v$ labels the 0-cells, and $p$ labels the 2-cells.  It is important to emphasize that we place an independent $d=2$ toric code model on each leaf, decoupled from all the other leaves.  Each 1-cell is contained in two leaves, so we place two qubits on every 1-cell, each associated with one of the two leaves containing the 1-cell.  Pauli operators can thus be labeled by a pair $(\ell, f)$ of a 1-cell $\ell$ and a leaf $f$, and for Pauli operators we write $Z^f_\ell$ and $X^f_\ell$.

We thus obtain a system of three stacks of decoupled $d=2$ toric code layers, with Hamiltonian
\begin{equation}
H_{{\rm stack}} = \sum_{f} H_{TC}(f) \text{,}  \label{eqn:hstack}
\end{equation}
where $f$ labels the leaves of the foliation structure.  To take the thermodynamic limit, we send the number of discretized leaves in each of the three foliations to infinity.

References~\onlinecite{ma17coupled,vijay17coupled} started with $H_{{\rm stack}}$ on the simple cubic lattice, and then added a certain term coupling the layers.  Here, rather than simply adding this coupling term, we will impose a $\zz$ 1-form symmetry, and then the same coupling will naturally arise as the most local symmetry-allowed term.  

The toric code $m$ particle excitations (magnetic flux excitations, in gauge theory language) reside on 2-cells $p$ with $B_p = -1$.  We pass to a dual cell structure, where each $n$-cell is replaced by dual a $3 - n$ cell intersecting it transversally.  For the standard foliation structure on $T^3$, the dual structure is the usual dual simple cubic lattice.  Viewing $m$ particles as residing on the dual 1-cells, we can view a configuration of $m$-particle excitations as a $\zz$-valued vector field on the dual cell structure.  Put another way, to each $m$-particle configuration can be associated a 1-manifold-with-boundary $M^1$, by taking the union of dual 1-cells occupied by $m$-particles.  It is thus natural to impose a $\zz$ 1-form symmetry where $m$-particle configurations are the charged objects.  We thus refer to $m$-particle configurations as $m$-strings.  More generally such strings are referred to in this context as ``$p$-strings,'' where the $p$ stands for ``particle,'' because such a string is made up of many point-like particle excitations.

It is important to note a change in perspective from the discussion in Sec.~\ref{subsec:hfs-charged}, where for a $q$-form symmetry we discussed charged \emph{operators} supported on $q$-dimensional submanifolds with boundary.  Here, we are considering excited states, where the configuration of excitations above a ground state is supported on a $q$-dimensional submanifold with boundary (in this case, $q=1$).  This is not the same thing if the excitations are not locally createable, which is of course the case for a toric code $m$ particle.  Below in Sec.~\ref{sec:charged} we discuss the consequences for the charged operators.

We use the original (\emph{i.e.} not dual) cell structure on $M$ to introduce $\zz$ 1-form symmetry operators following Sec.~\ref{subsec:cellular}.  In particular, symmetry operators are defined on 2-cycles $c_2 \in Z_2(M, \zz)$; this includes, as a special case, closed 2-manifolds $M^2 \subset M$ formed  by gluing together 2-cells.  The symmetry operator on $c_2$ is given by
\begin{equation}
U(c_2) = \prod_{p \in c_2} B_p \text{,}
\end{equation}
where $B_p$ is the $d=2$ toric code plaquette operator at $p$ (in the unique leaf containing $p$).  

This 1-form symmetry is easily seen to be non-faithful; in particular, $U(c_2) = 1$ whenever $c_2$ is a leaf or a sum of leaves.  We thus observe that the 1-form symmetry ``knows'' about the foliation structure.  In fact, we expect that the foliation structure can be deduced if one knows the symmetry operators $U(c_2)$ as a function of 2-cycles $c_2$.  We thus refer to this symmetry as a \emph{foliated 1-form symmetry}.  This is further justified by noting that, in addition to being non-faithful, this symmetry is also not topological, in that there exist 3-chains $c_3$ for which $U(\partial c_3) \neq 1$.  For example, we can take $c_3$ to be a single 3-cell, in which case we have
\begin{equation}
U(\partial c_3) = \prod_{p \in \partial c_3} B_p = \prod_{\ell \in  \partial c_3} X^{f_1(\ell)}_\ell X^{f_2(\ell)}_\ell \equiv {\cal B}_{c_3} \text{.} \label{eqn:small-symm-op}
\end{equation}
The second product is over the 1-cells $\ell$ contained in the boundary of the three-cell $c_3$, with $f_1(\ell)$ and $f_2(\ell)$ the two leaves containing $\ell$.  The resulting operator ${\cal B}_{c_3}$ is not the identity, and plays an important role below.

Imposing the 1-form symmetry means that terms in the Hamiltonian can only create closed $m$-strings, or $m$-loops.  The most local such term is precisely 
\begin{equation}
H_{{\rm coupling}} = - J_z \sum_\ell Z^{f_1(\ell)}_\ell Z^{f_2(\ell)}_\ell \text{,}
\end{equation}
the same coupling chosen in Refs.~\onlinecite{ma17coupled,vijay17coupled}, where again $f_1(\ell)$ and $f_2(\ell)$ are the two leaves containing $\ell$.   We emphasize that in the earlier works, the coupling was chosen not with any symmetry principle in mind, but simply as a means to realize the X-cube fracton model in the large-$J_z$ limit.  Here we observe that this coupling arises naturally upon imposing the $\zz$ 1-form symmetry.

At this point, upon increasing $J_z$, we expect $m$-loops to proliferate and condense as studied in Refs.~\onlinecite{ma17coupled,vijay17coupled}, resulting in the X-cube fracton phase.  As occurs on the simple cubic lattice,\cite{ma17coupled, vijay17coupled} it is likely that the foliated X-cube Hamiltonian of Eq.~(\ref{eqn:foliated-Xcube}), with modified coefficients, emerges in degenerate perturbation theory in the $J_z \to \infty$ limit, with the Pauli $Z$ terms arising at first order, and the Pauli $X$ terms arising at orders that will depend on the details of the foliation structure.  Unlike on the simple cubic lattice, there may be corrections to the foliated X-cube Hamiltonian, which could even affect the large $J_z$ ground state.  However, for our present purposes, the more important point is that the foliated X-cube fracton phase can be understood in terms of condensation of $m$-loops, just as on the simple cubic lattice.

A solvable Hamiltonian for the large-$J_z$ X-cube phase is
\begin{eqnarray}
\tilde{H}_{XC} &=& - J_z \sum_\ell Z^{f_1(\ell)}_\ell Z^{f_2(\ell)}_\ell \nonumber  \\ 
&+& H_{XC}(Z_\ell \to Z^{{\rm eff}}_\ell, X_\ell \to X^{{\rm eff}}_\ell ) \text{.}  \label{eqn:foliated-XC2}
\end{eqnarray}
The second term is the foliated X-cube Hamiltonian of Eq.~(\ref{eqn:foliated-Xcube}), but with Pauli operators replaced with new effective Pauli operators $Z^{{\rm eff}}_\ell$ and $X^{{\rm eff}}_\ell$ that we now describe.  The first term splits the four states on each 1-cell $\ell$ into two doublets, with $Z^{f_1(\ell)}_\ell Z^{f_2(\ell)}_\ell = 1$ in the ground state doublet.  The effective Pauli operators are defined by $X^{{\rm eff}}_\ell = X^{f_1(\ell)}_\ell X^{f_2(\ell)}_\ell$ and $Z^{{\rm eff}}_\ell = Z^f_\ell$; these act as $X$ and $Z$ Pauli operators within the ground-state doublet.  Note that the choice of leaf $f$ in the definition of $Z^{{\rm eff}}_\ell$ is arbitrary, because $Z^{f_1(\ell)}_\ell  =Z^{f_2(\ell)}_\ell$ within the ground-state doublet.  The  replacement $X_\ell \to X^{{\rm eff}}_\ell$ in  Eq.~\ref{eqn:foliated-XC2}   amounts to changing $B_{c_3} \to {\cal B}_{c_3}$ in the foliated X-cube Hamiltonian, where ${\cal B}_{c_3} = U(\partial c_3)$ is defined in Eq.~\ref{eqn:small-symm-op}.

The Hamiltonian of Eq.~\ref{eqn:foliated-XC2} is a sum of commuting terms, and is thus in the same phase (large-$J_z$ phase) for any finite $J_z > 0$.  Within the ground state subspace of the $J_z$ term, the Hamiltonian reduces by design to the foliated X-cube model.  In the context of the coupled-layer construction, we also refer to Eq.~(\ref{eqn:foliated-XC2}) as the foliated X-cube model.

\subsection{Charged operators}
\label{sec:charged}

We now turn to the charged operators, some understanding of which will be important for our discussion of breaking the 1-form symmetry below in Sec.~\ref{sec:breaking}.  The charged operators are somewhat subtle as compared to the more familiar cases of faithful or topological 1-form symmetries, because the one-dimensional charged objects are not locally createable, and the charged operators are thus not supported on one-dimensional regions.

\begin{figure}
        \includegraphics[width=\columnwidth]{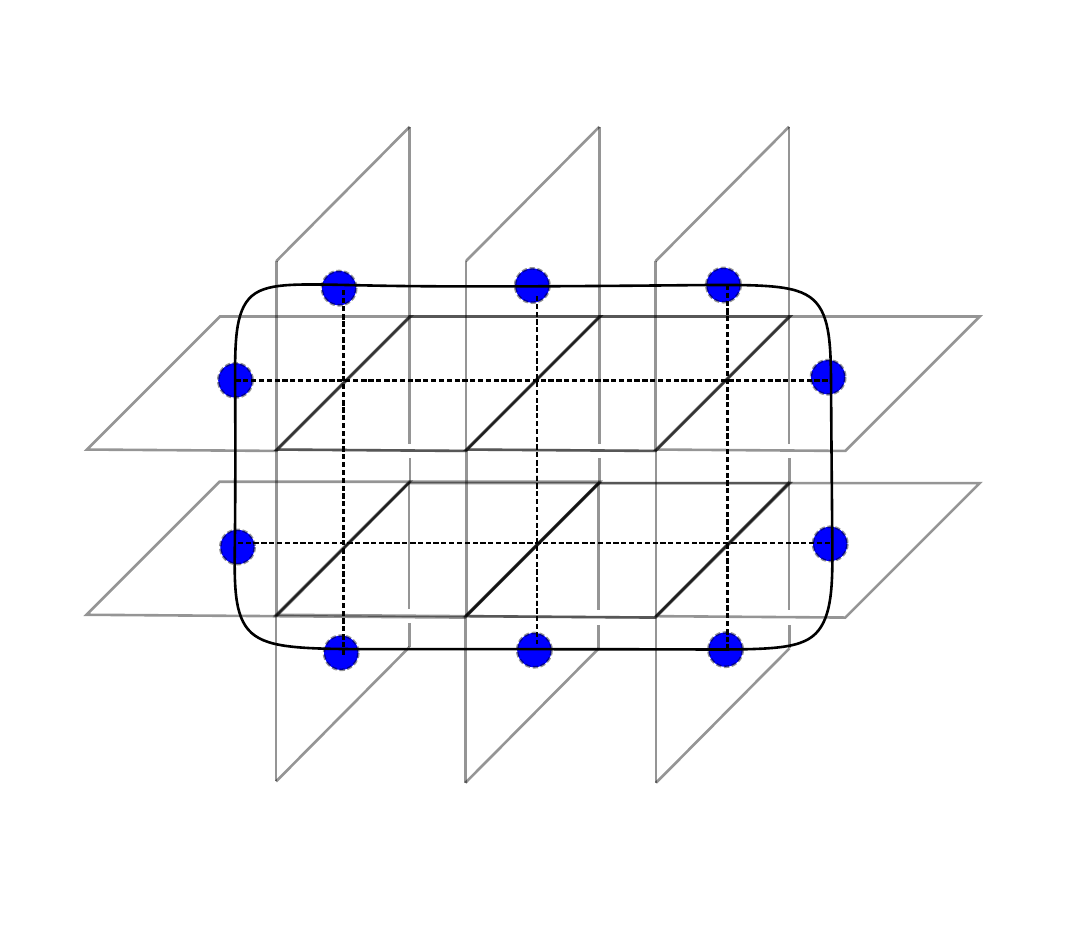}
        \caption{Illustration of a charged operator ${\cal O}(M^1)$. The black curve and blue dots correspond to the location of $M^1$ and $m$-particles respectively. The operator is formed by taking the product of line operators (dashed lines) within leaves. The choice of ${\cal O}(M^1)$ is not unique, as the line operators can be deformed within each leaf or may include closed $m$-line operators.  }
        \label{fig:mstring}
\end{figure}

Let $M^1$ be a 1-manifold, possibly with boundary, obtained by gluing together dual 1-cells.  We suppose $M^1$ intersects each leaf an even number of times, so that it is possible to create an $m$-string on $M^1$ without creating additional excitations.  This restriction on $M^1$ comes from the fact that each leaf can only accommodate even numbers of $m$-particle excitations.  A charged operator ${\cal O}(M^1)$ can be constructed by, within each leaf $f$, joining the plaquettes where $M^1$ intersects the leaf  with $m$-line operators (products of $Z^f_\ell$ on a dual curve joining the plaquettes), and then taking a product over leaves.  See Fig.~\ref{fig:mstring} for an example of such a charged operator.  There is some arbitrariness in the choice of line operators, so there is not a unique prescription to define ${\cal O}(M^1)$.  For instance, even if $M^1$ is empty, ${\cal O}(M^1)$ can be an arbitrary product of closed $m$-line operators.  However, the transformations of ${\cal O}(M^1)$ under 1-form symmetry only depend on $M^1$. It is important to emphasize that the support of ${\cal O}(M^1)$ is not $M^1$, and this operator  cannot be viewed as a one-dimensional object.

In the case where $M^1$ is the boundary of a disc $\Sigma$ (obtained by gluing dual 2-cells), there is a simple formula for ${\cal O}(M^1)$, namely
\begin{equation}
{\cal O}(M^1) = {\cal O}_{{\rm disc}}(\Sigma) = \prod_{\bar{p} \in \Sigma} Z^{f_1(\bar{p})}_{\bar{p}} Z^{f_2(\bar{p})}_{\bar{p}} \text{,}
\end{equation}
where $\bar{p}$ labels dual 2-cells, which are identified with 1-cells $\ell$.

In general we have
\begin{equation}
U(M^2) {\cal O}(M^1) = \Lambda {\cal O}(M^1) U(M^2) \text{,}
\end{equation}
with $\Lambda = (-1)^{\operatorname{Int}(M^2, M^1)}$, where $\operatorname{Int}(M^2, M^1)$ is the number of intersections between $M^2$ and $M^1$.  This relation is the analog of Eq.~\ref{eqn:qform-transformation}.

\subsection{Breaking of the foliated 1-form symmetry}
\label{sec:breaking}

Now we address the breaking of 1-form symmetry, considering first the standard foliation structure on $T^3$ and its simple cubic lattice discretization.  We expect that the small-$J_z$ phase with uncondensed $m$-loops (weakly coupled toric code layers) should have unbroken 1-form symmetry, while in the $m$-loop condensed X-cube phase, the symmetry is broken.  A simple way to see this would be to observe that the symmetry operators $U(c_2)$ act non-trivially in the ground state subspace in the large-$J_z$ phase.  

However, a subtlety arises:  every $U(c_2)$ projects to the identity operator in the ground state subspace of both phases.  This can be seen by studying the solvable points within the small- and large-$J_z$ phases provided by $H_{{\rm stack}}$ [Eq.~(\ref{eqn:hstack})] and $\tilde{H}_{XC}$ [Eq.~(\ref{eqn:foliated-XC2})], respectively.  Considering first $H_{{\rm stack}}$, $B_p = 1$ within the ground state subspace, so clearly also $U(c_2) = 1$ for any $c_2$.  This conclusion clearly holds independent of the choice of foliated spatial manifold $M$.

Next we consider the large-$J_z$ phase and $\tilde{H}_{XC}$.  Given $c_2 \in Z_2(M, \zz)$, let $[c_2] \in H_2(M, \zz)$ be the homology class.  If $[c_2] = 0$, then $c_2$ is a boundary, and $U(c_2)$ is a product of ${\cal B}_{c_3}$ operators for individual 3-cells, which all project to $1$ in the ground state subspace.  On the other hand if $[c_2]$ is non-trivial, then  $c_2 = f + \partial c_3$, \emph{i.e.} $c_2$ is the sum of a leaf $f$ and a boundary, because the leaves of the standard foliation structure generate $H_2(M, \zz)$.  Since $U(f) = 1$ (even away from the ground state subspace), we again have $U(c_2) = 1$ acting on ground states.  Unlike in the small-$J_z$ phase, this argument makes use of the choice of foliated spatial manifold.

It is important to emphasize that this subtlety does \emph{not} imply the 1-form symmetry remains unbroken in the X-cube phase.  Indeed, the same phenomenon occurs in the $d=2$ toric code model on $S^2$, where the ground state is unique even in the topologically ordered 1-form symmetry-breaking phase.  There, of course, by working on $T^2$ instead of $S^2$, we find a non-trivial action of 1-form symmetry operators in the ground state subspace in the symmetry-breaking phase.

This suggests that we should seek a different foliated spatial manifold $M$, where $U(c_2)$ does not always project to the identity operator in the ground state subspace of the X-cube phase.  Examining the above argument, the key fact about the standard foliation structure on $T^3$ was that the foliation leaves generate $H_2(M, \zz)$; the argument goes through for any foliated manifold where this is true.  Therefore, we need to find a foliated manifold where $H_2(M, \zz)$ is not generated by the leaves.  Put another way, there must exist a 2-cycle with non-trivial homology class, that is not homologous to a leaf or sum of leaves.

With this requirement in mind, we consider $T^3$ with the twisted foliation structure shown in Fig.~\ref{fig:foliations}(b).  As for the standard foliation structure, it is convenient to view $T^3$ as a $L \times L \times L$ cube with opposite faces identified, where here we take $L$ to be an even integer. We choose two of the three foliations to be the same as in the standard foliation structure, with $yz$- and $xz$-plane leaves.  The leaves of the third foliation, which we refer to as $\widetilde{xy}$ leaves, are $xy$-planes tilted along the $y$-direction as shown, so that upon traversing a  $\widetilde{xy}$ leaf along the $y$-direction from $y = 0$ to $y = L$, one moves $+L/2$ in the $z$-direction.  We note that of course there are many similar twisted foliation structures, but considering this specific one will be enough for our purposes, and so we refer to ``the'' twisted foliation structure.

To study the foliated X-cube model Eq.~(\ref{eqn:foliated-XC2}), we need to choose a finite subset of leaves.  We choose $yz$ leaves with unit spacing along the $x$-direction, $xz$ leaves with unit spacing along $y$, and $\widetilde{xy}$ leaves with unit spacing along $z$.  Thus we have $L$ distinct $yz$ leaves, $L$ $xz$ leaves, and $L/2$ $\widetilde{xy}$ leaves.  The resulting model can equivalently be obtained by starting with the X-cube model on a $L \times L \times L$ simple cubic lattice, and twisting the boundary conditions using the lattice translation symmetry so that
 \begin{eqnarray}
 (T_x)^L &=& (T^z)^L = 1 \\
 (T_y)^L &=& (T^z)^{L/2} \text{,}
 \end{eqnarray}
 where $T_x$, $T_y$ and $T_z$ are translation by one lattice constant in the $x$, $y$ and $z$ directions.  We note that these boundary conditions are a special case of more general boundary conditions for the X-cube model studied in Ref.~\onlinecite{rudelius20toappear}.

\begin{figure}
	\includegraphics[width=\columnwidth]{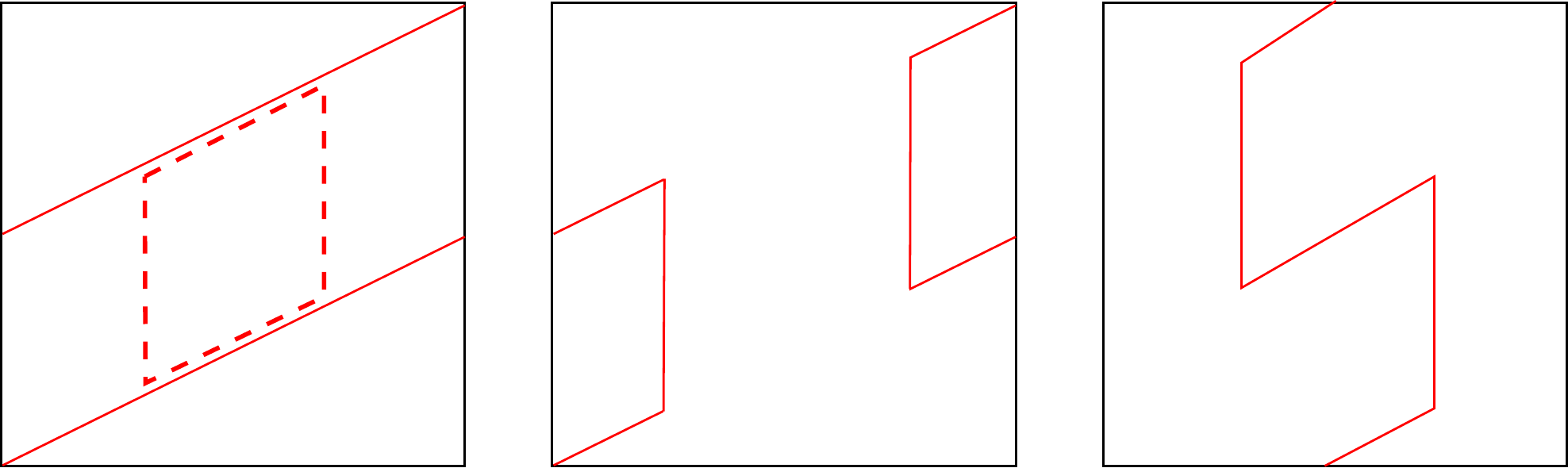}
	\caption{Graphical demonstration that a $\widetilde{xy}$ leaf (solid line, left panel) in the twisted foliation structure on $T^3$ is homologous to a $xz$ leaf (in homology with $\zz$ coefficients).  In each panel a $yz$-plane cross section of $T^3$ is shown.  First, one adds the boundary of a cylinder running along the $x$-direction, shown by the dashed line in the left panel.  Joining this with the $\widetilde{xy}$ leaf results in the cycle shown in the middle panel, which can then be slid to the left or right to obtain cycle in the right panel, which is clearly homologous to a $xz$ leaf.
	   \label{fig:leaf-homology}}
\end{figure}

\begin{figure}
	\includegraphics[width=0.7\columnwidth]{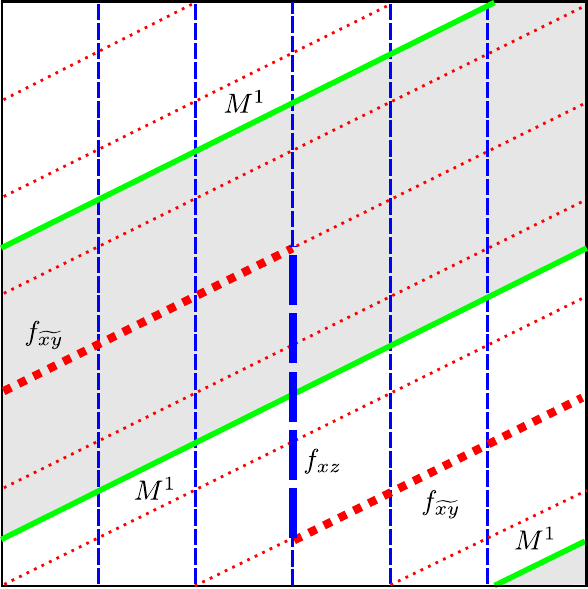}
	\caption{Illustration of a symmetry operator $U(c^{xy}_2)$ that acts non-trivially in the ground state space of the X-cube phase on $T^3$ with twisted foliation structure, and a charged operator ${\cal O}(M^1)$ that anticommutes with $U(M^{xy}_2)$.  A $yz$-plane cross section of $T^3$ is shown, with the dotted (red online) and dashed (blue online) lines indicating $\widetilde{xy}$ and $xz$ leaves as in Fig.~\ref{fig:foliations}(b).  $c_2^{xy}$ is a union of regions from the $f_{\widetilde{xy}}$ and $f_{xz}$ leaves, and the cross section of $c^{xy}_2$ is indicated by the thick dotted and dashed lines.  The symmetry operator ${\cal O}(M^1)$ creates a $m$-loop on $M^1$, which lies in a fixed $yz$-plane, and is indicated by the solid thick diagonal line (green online).  ${\cal O}(M^1)$ is a product of $Z^f_\ell$ Pauli operators over $\Sigma_{yz}$, which is a surface lying in the same $yz$-plane as $M^1$, shown in the figure by the shaded region (see text for further details).	
	  \label{fig:symm-op}}
\end{figure}

Now recall that $H_2(T^3, \zz) = \zz^3$, which is generated by one each of $xy$, $yz$ and $xz$ planes.  For the twisted foliation structure, an $\widetilde{xy}$ leaf is homologous to an $xz$ leaf, as illustrated in Fig.~\ref{fig:leaf-homology}, so the leaves of the foliation only generate a $\zz^2$ subgroup of $H_2(T^3, \zz)$.  

We take $c_2 = c^{xy}_2$ as shown in Fig.~\ref{fig:symm-op}, which consists of regions from two leaves $f_{\widetilde{xy}}$ and $f_{xz}$, intersecting along lines ${\cal L}_1$ and ${\cal L}_2$.  $c^{xy}_2$ generates the remaining $\zz$ factor in $H_2(T^3, \zz)$, and is thus not homologous to a leaf or sum of leaves.  The symmetry operator $U(c^{xy}_2)$ reduces to a product along the lines ${\cal L}_1, {\cal L}_2$:
\begin{equation}
U(c^{xy}_2) = \prod_{\ell \in {\cal L}_1} X^{f_1(\ell)}_\ell X^{f_2(\ell)}_\ell   \prod_{\ell \in {\cal L}_2} X^{f_1(\ell)}_\ell X^{f_2(\ell)}_\ell  \text{.} \label{eqn:twisted-symm-op}
\end{equation}

Now again considering the large-$J_z$ phase at its solvable point with Hamiltonian $\tilde{H}_{XC}$, we show that $U(c^{xy}_2)$ indeed acts non-trivially within the ground state subspace.  We do this by exhibiting a charged operator that commutes with $\tilde{H}_{XC}$ and anticommutes with $U(c^{xy}_2)$.  That is, we find an anticommuting pair of logical operators, one of which is $U(c^{xy}_2)$.  The operator we need is ${\cal O}(M^1) = \prod_{\bar{p} \in \Sigma_{yz}} Z^{f}_{\bar{p}}$, where $M^1$ is as shown in Fig.~\ref{fig:symm-op}, the leaves $f$ are all chosen to be $xz$ leaves, and $\Sigma_{yz}$ is a parallelogram within a fixed $yz$-plane.  It is evident from Fig.~\ref{fig:symm-op} that $M^1$ intersects $c^{xy}_2$ exactly once, so $U(c^{xy}_2)$ and ${\cal O}(M^1)$ anticommute.  Moreover, because $M^1$ is closed, ${\cal O}(M^1)$ commutes with $\tilde{H}_{XC}$.  It follows that there are ground states with both $+1$ and $-1$ eigenvalues of $U(c^{xy}_2)$, \emph{i.e.} the symmetry operator acts non-trivially in the ground state subspace.  

While the behavior of $U(c^{xy}_2)$ projected to the ground state subspace was obtained by considering special solvable Hamiltonians, it is robust within the two phases as long as the 1-form symmetry is maintained, because $U(c^{xy}_2)$ commutes with the Hamiltonian.  This is obvious in small-$J_z$ phase, where all ground states have eigenvalue $+1$ under $U(c^{xy}_2)$.  In the X-cube phase, one might worry that the ground states with eigenvalues $+1$ and $-1$ could split in energy upon adding some perturbation.  However, this cannot happen because the local density matrices of these states are the same, a conclusion that follows from the topological order of the X-cube model on the simple cubic lattice,\cite{vijay16fracton} a model that only differs from the X-cube model on the twisted foliation of $T^3$ by a different choice of periodic boundary conditions.

\subsection{Variants of this construction}
\label{subsec:variants}

To recap our discussion so far, we (1) started with layers of $d=2$ toric code models placed on the leaves of a discretized foliation structure, (2) imposed a 1-form symmetry that turned out to depend on the foliation structure, and (3) condensed charged loops to obtain the X-cube fracton phase via breaking of the foliated 1-form symmetry.  In this section, we discuss variations on this procedure with the same final result.  In particular, by making different choices, we can replace steps that involve condensation with gauging of certain symmetries.

To illustrate the possibilities, we begin with a trivially gapped system with a \emph{faithful} $\zz$ 1-form symmetry.  To keep the discussion concrete, we introduce a model defined by placing a single qubit on each link $\ell$ of the $d=3$ simple cubic lattice, and with Hamiltonian
\begin{equation}
H = - \sum_\ell X_\ell \text{.}
\end{equation}
Links $\ell$ are the same as plaquettes $\bar{p}$ of the dual lattice, so we can write $X_{\bar{p}} \equiv X_\ell$. We introduce a cell structure using the dual cubic lattice, and 1-form symmetry operators are then defined on cellular 2-cycles $c_2 \in Z_2(M, \zz)$,
\begin{equation}
U(c_2) = \prod_{\bar{p} \in c_2} X_{\bar{p}} \text{.}
\end{equation}
This is easily seen to be a faithful 1-form symmetry.

Thus far no foliation structure is explicitly involved; while our discussion focuses on the cubic lattice, the constructions above can be made for any cell structure on $M$.  To proceed we introduce a foliation structure on $M$ as before, choosing for concreteness $M = T^3$ and taking the standard foliation structure.  Now, for each leaf $f$, the symmetry operator $U(f)$ can be viewed as generating a $\zz$ 0-form symmetry supported on the leaf.  We gauge all of these $\zz$ symmetries, giving a deconfined $d=2$ $\zz$ gauge theory coupled to $\zz$ matter on each leaf.  We thus obtain a stack of decoupled $d=2$ $\zz$ gauge theories, each of which is equivalent to a toric code model.  

After this gauging step, we are still left with a residual 1-form symmetry, which is now the same foliated 1-form symmetry discussed above.  This is so because $U(f)$ measures the total $\zz$ gauge charge on each leaf of the foliation, which must vanish, so $U(f) = 1$.  Here, the charged objects under the residual symmetry are $e$-strings; that is, they are strings made up of $\zz$ gauge charges (toric code $e$ particles).  This is not an important difference, because of course $d=2$ $\zz$ gauge theory enjoys an electric-magnetic duality that exchanges $e$ and $m$ excitations.  Indeed, the construction we are describing is the electric-magnetic dual of the original coupled-layer construction.  At this point, we can use the residual 1-form symmetry to add symmetry-respecting coupling terms between the $d=2$ layers, and to condense charged loops.  This again results in the X-cube fracton phase.

Another variation of this construction is possible, where again we start with a trivially gapped system with faithful $\zz$ 1-form symmetry, and first condense the charged loops to obtain a $d=3$ deconfined $\zz$ gauge theory ($d=3$ toric code).  Then we introduce the same foliation structure, and gauge the same collection of $\zz$ 0-form symmetries as above.  Such a construction was in fact already considered in Ref.~\onlinecite{williamson19SETfracton} and shown to result in the X-cube model.

\subsection{Relation to emergent infrared symmetries of the X-cube phase}
\label{sec:XCIR}

A natural question is how the foliated 1-form symmetry is related to emergent IR symmetries within the X-cube phase.  To motivate the discussion of this section, we first review an analogous relationship in the $d=2$ toric code model.  We consider the faithful 1-form $Z$-symmetry of the toric code, which is a UV symmetry.  This symmetry is spontaneously broken in the small-$h$ topologically ordered phase.  Within this phase, below the gap to vertex excitations (electric charges in gauge theory language), there is also an IR topological 1-form symmetry, which can be obtained by projecting the $Z$-symmetry into the ground state subspace.  This IR symmetry is emergent, in the sense that it is robust even if the $Z$-symmetry is explicitly broken at the microscopic level.

Here we discuss a similar but more subtle relationship that holds in the X-cube fracton phase, between the UV foliated 1-form symmetry and an IR generalized symmetry, which appears not to be a $q$-form symmetry at all.  Our intent is only to illustrate this relationship and not to give a complete discussion; in particular, we do not fully analyze the emergent IR symmetries of the X-cube phase. We refer the reader to Ref.~\onlinecite{seiberg20exotic3dZN} for a discussion of the symmetries of a continuum field description of the X-cube model and its $\z_n$ generalization.

We focus for simplicity on the cubic lattice (equivalently, on the discretized standard foliation structure on $T^3$), and  consider the model $\tilde{H}_{XC}$ of Eq.~(\ref{eqn:foliated-XC2}).  It is well-known and easy to check that $\tilde{H}_{XC}$ commutes with string logical operators of the form
$S_{\cal L} = \prod_{\ell \in {\cal L}} X^{f_1(\ell)}_\ell  X^{f_2(\ell)}_\ell$, where ${\cal L}$ is any closed straight line path of 1-cells.\cite{vijay16fracton}  If we restrict to the ground state subspace of the $J_z$ terms in $\tilde{H}_{XC}$, then $X_\ell \equiv X^{f_1(\ell)}_\ell  X^{f_2(\ell)}_\ell$, and the $S_{\cal L}$ become familiar string logical operators of the X-cube model.  

The ${\cal S}_{\cal L}$ operators can be viewed as symmetry operators of a generalized symmetry, which we refer to as ${\cal S}$-symmetry. This symmetry looks similar to a 2-form symmetry;  however, going along with the restricted mobility of lineon excitations and differing from a 2-form symmetry, the line ${\cal L}$ cannot be bent if one wants to preserve $[S_{\cal L}, \tilde{H}_{XC}]=0$.  Nonetheless, the ${\cal S}$-symmetry is broken in the X-cube phase, as the ${\cal S}_{\cal L}$ operators act non-trivially in the ground state subspace.  

So far the ${\cal S}$-symmetry is a UV symmetry, but we can project it to the subspace of states below the gap to fracton excitations, upon which it becomes an emergent IR symmetry ${\cal S}_{IR}$.  This symmetry is robust below the fracton gap, even if we explicitly break the UV ${\cal S}$-symmetry.  To understand why, it is helpful to recall that in conventional $d=2$ $\zz$ gauge theory, below the gap to electric charge excitations one can describe the system by an effective pure $\zz$ gauge theory, which has a topological 1-form symmetry.  In terms of the original variables at the UV scale, the symmetry operators of this topological 1-form symmetry are ``fattened'' strings, where the fattening is a consequence of the non-trivial relationship between microscopic and effective degrees of freedom.\cite{hastings05quasiadiabatic}

Coming back to the X-cube model, below the fracton gap, one has an effective generalized gauge theory with a local constraint ${\cal B}_{c_3} = 1$.  This model has the ${\cal S}$-symmetry, by an argument given below.  Therefore the original X-cube model will have the IR symmetry ${\cal S}_{IR}$ at scales below the fracton gap, with ${\cal S}_{\cal L}$ operators replaced by fattened lineon strings.

It remains to understand how the ${\cal S}$-symmetry is related to the foliated 1-form symmetry.  First of all, the operators ${\cal S}_{\cal L}$ are not part of the foliated 1-form symmetry, because there is no way to write ${\cal S}_{\cal L}$ as a product of $B_p$ operators.  Instead, we claim that any local Hamiltonian $H$ invariant under the foliated 1-form symmetry also necessarily enjoys the ${\cal S}$-symmetry.  Suppose $\Gamma \in C_3(T^3, \zz)$ is a $k \times k \times k$ cube within the cubic lattice, represented as a sum of elementary cubes $c_3 \in \Gamma$. We consider the 1-form symmetry operator
\begin{equation}
U(\partial \Gamma) = \prod_{c_3 \in \Gamma} {\cal B}_{c_3} \text{.}
\end{equation}
This operator is a product of $X^{f_1(\ell)}_\ell X^{f_2(\ell)}_\ell$ over the edges of the large cube $\Gamma$.  Supposing that $k$ is much larger than the finite range of terms appearing in $H$, if we zoom in to the center of one of the edges of $\Gamma$, we conclude that the Hamiltonian must commute with a product of $X^{f_1(\ell)}_\ell X^{f_2(\ell)}_\ell$ along a straight line.  In particular, $H$ must commute with ${\cal S}_{\cal L}$, and we have the ${\cal S}$-symmetry.  The same argument shows that the generalized gauge theory with constraint ${\cal B}_{c_3} = 1$ also has the ${\cal S}$-symmetry; there $U(\partial \Gamma)$ must commute with $H$ because $U(\partial \Gamma) = 1$.

To summarize, we argued that a local Hamiltonian with foliated 1-form symmetry necessarily also has the ${\cal S}$-symmetry, even though the ${\cal S}$-symmetry is not part of the foliated 1-form symmetry.  The ${\cal S}$-symmetry then descends at low energy -- below the fracton gap -- to an emergent IR symmetry of the X-cube phase.  Such a situation is not special to fracton phases; indeed, analogous statements can be made about the $d=2$ toric code model.  There, suppose we consider a 1-form-like symmetry with the same symmetry operators $U(c_1)$ as for the $Z$-symmetry, but now where $c_1 \in B_1(M, \zz)$, \emph{i.e.} we only consider symmetry operators on 1-boundaries.  For a local Hamiltonian, by arguments paralleling those above, the presence of this symmetry implies the full $Z$ 1-form symmetry, which then descends to an IR topological 1-form symmetry in the topologically ordered toric code phase.

\section{Generalized $p$-string and gauging constructions}
\label{sec:generalized}

In the previous section, we described the $p$-string condensation route to the X-cube fracton phase in terms of breaking of a foliated 1-form symmetry.  We also argued in Sec.~\ref{subsec:variants} that certain constructions based on gauging symmetries are essentially the same as $p$-string condensation.  Here, we take a step back from the specific example of the X-cube fracton order, and give a schematic description of more general $p$-string condensation mechanisms from a higher-form symmetry point of view.  In Sec.~\ref{sec:u1} we will see that the $p$-string condensation mechanism for the rank-2 ${\rm U}(1)$ scalar charge theory,\cite{radzihovsky20vector} as well as a related construction based on gauging symmetry,\cite{williamson19SETfracton} is encompassed within this point of view.

We describe  $p$-string condensation as a three-step procedure.  In step (1), we begin with some collection of decoupled systems $S_\alpha$ labeled by $\alpha$.  We choose the $S_\alpha$ to be ``ordinary'' topological phases or gauge theories without fractons.  It  is important that the $S_\alpha$ should have fractional (\emph{i.e.} not locally createable) particle excitations.  Depending on the spatial dimension $d$ and the nature of the $S_\alpha$ systems, we may already need to introduce some  geometrical structure at this stage.  For instance, in the coupled-layer construction of the $X$-cube model, the $S_\alpha$ are layers of $d=2$ toric code, and placing these layers can be accomplished with a foliation structure.

Next, step (2) imposes a 1-form symmetry such that the charged objects are configurations of certain particle excitations of the $S_\alpha$'s.  Some such configurations can thus be viewed as extended one-dimensional charged objects, and these are the $p$-strings.  Without some geometrical structure, it is not natural to view point-like particles as forming the charged objects of a 1-form symmetry, so this step either relies on the geometrical structure introduced in (1), or relies on some geometry that must be introduced at this stage.  In the X-cube model example, the relevant geometry is the foliation structure already introduced at step (1).  In general we  can expect this 1-form symmetry to be non-faithful due to global constraints on how many particles can exist within each system $S_\alpha$, arising from the fractional nature of the particle excitations.  In the X-cube model, the non-faithfulness arose from the fact that each toric code layer contains an even number of $m$ particles.  

It is natural to expect that imposing the 1-form symmetry allows for local creation of small closed $p$-strings.  Increasing the corresponding coupling strength is then expected to lead to $p$-string condensation and spontaneous breaking of the 1-form symmetry -- this is step (3).  Whether this is actually possible may depend on choices made in step (1) and step (2).  For instance, it is not expected that $p$-strings built from Abelian anyons with non-trivial self statistics can condense.  Moreover, one can imagine making choices such that any bounded closed $p$-string carries some gauge charge and is thus not locally createable.  Finally, one needs to study the resulting $p$-string condensed phase to understand its properties, including its stability to perturbations, whether or not it supports fracton excitations, and so on.

As in Sec.~\ref{subsec:variants} above, it is possible to carry out these steps in a different order.  For instance, following that earlier discussion in this more abstract setting, we can begin with a trivial gapped state enjoying a faithful 1-form symmetry, and then gauge some 0-form subgroups of this symmetry, as a means of introducing the decoupled systems $S_\alpha$.  There is then a residual non-faithful 1-form symmetry that can be understood as a quotient of the original faithful 1-form symmetry by the gauged subgroup.  The gauge charges of the $S_{\alpha}$ are then charged under the residual 1-form symmetry, and thus form $p$-strings, whose condensation can be studied.  Alternatively, one can start from a phase where the faithful 1-form symmetry is already spontaneously broken, and then gauge the same 0-form subgroups.  This is expected to result immediately in the same $p$-string condensed phase, as the 1-form symmetry under which the $p$-strings are charged is already broken.

Various further generalizations are possible.  For instance, there is no obvious reason to restrict attention only to 1-form symmetries, and we can consider $q$-form symmetries with $q > 1$ and the corresponding condensation of $q$-dimensional $p$-membranes built from fractional particle excitations.  Indeed, Ref.~\onlinecite{ma17coupled} already studied a kind of condensation of two-dimensional $p$-membranes, which may be possible to understand from this perspective.

\section{Rank-2 ${\rm U}(1)$ scalar charge theory and framed 1-form symmetry}
\label{sec:u1}

Here we put the program of Sec.~\ref{sec:generalized} into practice, by considering an application to fractonic ${\rm U}(1)$ tensor gauge theories on the lattice.\cite{Xu1,Xu2,pretko17subdimensional}  In particular, we consider one of the simplest such models, the rank-2 ${\rm U}(1)$ scalar charge theory (simply ``scalar charge theory,'' for short).\cite{pretko17subdimensional}  Williamson, Bi and Cheng showed that this model can be obtained in $d=3$ starting from a single ${\rm U}(1)$ vector gauge theory, upon gauging certain global symmetries.\cite{williamson19SETfracton}  Recently, two of the authors (LR and MH) obtained the scalar charge theory in $d$ spatial dimensions from a different point of view, developing a generalization of $p$-string condensation.\cite{radzihovsky20vector}  The latter construction begins with $d + 1$ ${\rm U}(1)$ vector gauge theories linked via an unusual Gauss' law constraint.  For $d > 2$, when a certain coupling is larger than a critical value, $p$-string condensation occurs, and the scalar charge theory emerges as a low-energy effective description of the condensed phase.   Here, we develop these constructions from the viewpoint of Sec.~\ref{sec:generalized}, which also makes it clear that they are better thought of as two variants of the same construction.  We start with the gauging construction in Sec.~\ref{sec:u1-gauging}, and then proceed to give a brief account of the $p$-string condensation construction in Sec.~\ref{sec:u1-pstring}.

A product of this analysis is some insight into the spatial geometry needed to define the scalar charge theory, via a non-faithful 1-form symmetry.  For reasons that we elaborate on below, we dub this symmetry a \emph{framed} 1-form symmetry, to distinguish it from the foliated 1-form symmetry of Sec.~\ref{sec:xcube}.  It is temping to conclude that the framed 1-form symmetry is spontaneously broken in the deconfined phase of the scalar charge theory.  However, in contrast to the case of $p$-string condensation for the X-cube model, we do not yet have a precise characterization of framed 1-form symmetry-breaking, such as the non-trivial action of symmetry operators in the ground state subspace discussed in Sec.~\ref{sec:breaking}.  This point is discussed further below.

We begin by imposing some geometrical structure on the closed spatial $d$-manifold $M$.  Similar to the X-cube case of Sec.~\ref{sec:xcube}, we consider a $d$-multifoliation by $(d-1)$-dimensional compact leaves, except here we take $M$ and the leaves to be oriented, with the leaves inheriting their orientations from that of $M$.  Pairs of leaves are assumed to intersect transversally at $(d-2)$-submanifolds, triples of leaves at $(d-3)$-submanifolds, and so on.  We expect it is possible to allow for the multifoliation to be singular, but for simplicity we will not consider this possibility.

To visualize this structure, it is helpful to take $d=3$ and $M = T^3$, with an orientation of $T^3$ given by choosing a right-handed coordinate system.  We introduce a standard multifoliation structure on $T^3$, where the three foliations consist of $xy$, $yz$ and $xz$ planes.  The only difference from Sec.~\ref{sec:xcube} is that each plane now comes with a given orientation.  More generally, in $d$ dimensions we have a standard foliation structure on $T^d$, where each of the $d$ foliations consists of hyperplanes normal to one of the coordinate directions.

Again as in Sec.~\ref{sec:xcube}, we choose a non-empty finite set of leaves in each of the $d$ foliations to obtain a cell structure on $M$.  Labeling the $d$ foliations with an index $\alpha = 1,\dots,d$, we note that each $(d-1)$-cell $\ell_{d-1}$ is contained in a unique leaf, which is itself contained in the foliation with index $\alpha$.  We can therefore define a foliation index $\alpha(\ell_{d-1})$ associated to each $(d-1)$-cell.  Moreover, every $(d-1)$-cell comes with a given orientation inherited from the leaf containing it.  For the standard foliation structure on $T^d$, and taking equally spaced leaves in each of the foliations, the cell structure thus obtained is a $d$-dimensional hypercubic lattice.

\subsection{Gauging construction and framed 1-form symmetry}
\label{sec:u1-gauging}

We first describe the construction of the scalar charge theory in terms of gauging symmetry,\cite{williamson19SETfracton} from the point of view of Sec.~\ref{sec:generalized}.  This will allow us to introduce the framed 1-form symmetry.  On each $(d-1)$-cell $\ell_{d-1}$, we place a ${\rm U}(1)$ quantum rotor, with integer-valued number operator $e_{\alpha}(\ell_{d-1})$ and conjugate phase $a_{\alpha}(\ell_{d-1})$, where $\alpha = \alpha(\ell_{d-1})$.  Each $\ell_{d-1}$ is intersected transversally by a dual 1-cell $\bar{\ell}_1$, where $\bar{\ell}_1$ comes with a given orientation, so we can view $e_{\alpha}$ and $a_{\alpha}$ as lattice vector fields residing on dual 1-cells.  Indeed, we would like to view $e_{\alpha}$ and $a_{\alpha}$ as the electric field and vector potential of a compact ${\rm U}(1)$ gauge theory coupled to dynamical charged matter.  Rather than introduce a separate field for the matter degrees of freedom and impose a Gauss' law constraint, we equivalently leave $e_{\alpha}$ unconstrained.  Then $\rho \equiv \nabla \cdot e$, which is defined on dual 0-cells, is interpreted as the density of ${\rm U}(1)$ gauge charge.

\begin{figure}[h]
	\includegraphics[width=\columnwidth]{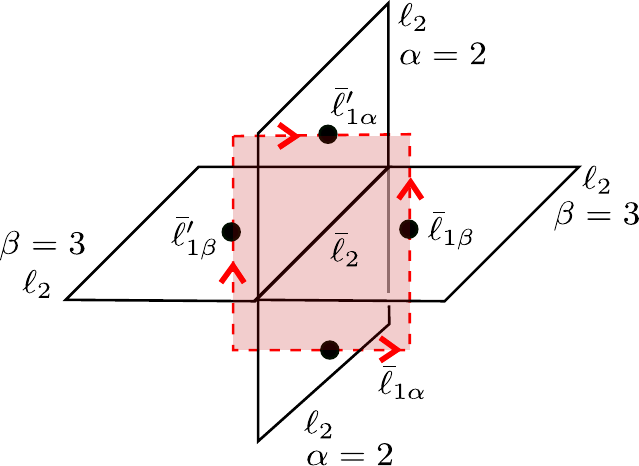}
	\caption{Illustration of the geometry involved in the definition of the magnetic field $b_{\alpha \beta}(\bar{\ell}_2)$ in three spatial dimensions.  Four 2-cells $\ell_2$, with foliation indices $\alpha = 2$ and $\beta = 3$, intersect at a single 1-cell (not labeled).  Each 2-cell is pierced (intersected transversally) by a dual 1-cell $\bar{\ell}_1$, indicated by dashed lines (red online) with the given orientations shown by the arrows.  The shaded region (red online) is a dual 2-cell $\bar{\ell}_2$, on which the magnetic field operator is defined and whose boundary consists of the four dual 1-cells as expressed in Eq.~(\ref{eqn:dual-2-cell-bdy}).}
		\label{fig:cells}
\end{figure}

This leads us to consider the Hamiltonian
\begin{equation}
H = U \sum_{\ell_{d-1}} [ e_{\alpha}(\ell_{d-1}) ]^2 - K \sum_{\bar{\ell}_2} \cos [ b_{\alpha \beta} (\bar{\ell}_2 ) ] \text{.}
\label{eqn:ungaugedU1}
\end{equation}
Here we have introduced the magnetic field $b_{\alpha \beta} (\bar{\ell}_2 )$, which is defined on dual 2-cells.  Each dual 2-cell $\bar{\ell}_2$ transversely intersects a unique $(d-2)$-cell $\ell_{d-2}$, which in turn lies in the intersection of two leaves with foliation indices $\alpha$ and $\beta$ ($\alpha \neq \beta$).  The boundary $\partial \bar{\ell}_2$ is a sum of four dual 1-cells,
\begin{equation}
\partial \bar{\ell}_2 = \bar{\ell}_{1 \alpha} + \bar{\ell}_{1 \beta} - \bar{\ell}'_{1 \alpha} - \bar{\ell}'_{1 \beta} \text{,} \label{eqn:dual-2-cell-bdy}
\end{equation}
as illustrated in Fig.~\ref{fig:cells} for the $d=3$ case.  Here, the foliation indices $\alpha$ and $\beta$ are those of the $(d-1)$-cells pierced by each dual 1-cell.  The magnetic field is given by
\begin{equation}
b_{\alpha \beta}(\bar{\ell}_2) = a_\alpha(\bar{\ell}_{1 \alpha}) + a_{\beta}(\bar{\ell}_{1 \beta}) 
- a_{\alpha}(\bar{\ell}'_{1 \alpha}) - a_{\beta}(\bar{\ell}'_{1 \beta}) \text{,}
\end{equation}
 a discrete line integral of $a_{\alpha}$ over the boundary $\partial \bar{\ell}_2$.
If $d \geq 3$, the model has two phases, with a confined phase at small $K / U$, and a deconfined phase at large $K / U$.

We now impose an electric ${\rm U}(1)$ 1-form symmetry.  We let $Z_{d-1}(M, \z)$ be the group of cellular $(d-1)$-cycles, whose elements $c_{d-1} \in Z_{d-1}(M, \z)$ can be written as sums $c_{d-1} = \sum_{\ell_{d-1}}  n(\ell_{d-1}) \ell_{d-1}$, with $n(\ell_{d-1}) \in \z$, and satisfying $\partial c_{d-1} = 0$.  The operators generating the 1-form symmetry are then 
\begin{equation}
\Phi(c_{d-1}) = \sum_{\ell_{d-1}} n(\ell_{d-1}) e_\alpha (\ell_{d-1}) \text{.}  \label{eqn:u1-symm-ops}
\end{equation}
This operator measures the electric flux through the cycle $c_{d-1}$.  It is important to note that this makes sense only because of the given orientation on each $\ell_{d-1}$; that is, $e_\alpha(\ell_{d-1})$ measures the electric flux through $\ell_{d-1}$, taking the given orientation. Because there are no constraints on $e_{\alpha}$, this is clearly a faithful 1-form symmetry.  As discussed in Sec.~\ref{subsec:hfs-general}, imposing this symmetry makes the dynamical charged matter completely immobile, \emph{i.e.} it forbids a kinetic energy term for the charged matter.  Put another way, while lines of electric field are free to fluctuate, open ends of field lines are immobile as long as the 1-form symmetry is imposed.

It is important to note that the faithful 1-form symmetry is indeed a symmetry of the Hamiltonian Eq.~(\ref{eqn:ungaugedU1}).  This is so because $e^{i b_{\alpha \beta} (\bar{\ell}_2 )}$ creates a small closed loop of electric field running along the boundary of $\bar{\ell}_2$.  Adding such a loop of electric field does not change the value of $\Phi(c_{d-1})$ for any cycle $c_{d-1}$, so $b_{\alpha \beta} (\bar{\ell}_2 )$ commutes with all the 1-form symmetry operators.

We now single out $d$ 0-form subgroups of $Z_{d-1}(M, \z)$ to be gauged, which will result in a residual non-faithful 1-form symmetry.  Generalizing from Ref.~\onlinecite{williamson19SETfracton} to the present context, we let $Q_\beta$ be the sum of all electric field operators pointing in the $\beta$-direction.  That is we choose
\begin{equation}
Q_\beta = \sum_{ \{ \ell_{d-1} | \beta = \alpha(\ell_{d-1}) \} } e_{\beta}(\ell_{d-1}) \text{.}  \label{eqn:Qbeta}
\end{equation}
This is a 1-form symmetry operator, obtained by taking $c_{d-1}$ to be a sum of all the leaves in the $\beta$-foliation.  We view each $Q_\beta$ as generating a 0-form ${\rm U}(1)$ symmetry, which arises as a subgroup of the faithful 1-form symmetry.

We gauge these $d$ 0-form symmetries, introducing for each one an integer-valued ${\rm U}(1)$ electric field $\bE_\beta$ and conjugate compact vector potential $\bA_\beta$.  To arrive at a lattice model, we note that $e_\beta$ is the gauge charge density of the $\beta$ gauge theory, so charges of this theory reside on $(d-1)$-cells with $\beta = \alpha(\ell_{d-1})$.  Therefore it is natural to place $\bE_\beta$ and $\bA_\beta$ on links joining pairs of such type-$\beta$ $(d-1)$-cells.  For instance, if we take the standard multifoliation on $T^d$ and resulting hypercubic lattice cell structure, the type-$\beta$ $(d-1)$-cells themselves form the vertices of a hypercubic lattice, on whose nearest-neighbor links we can place $\bE_\beta$ and $\bA_\beta$.  This is precisely the lattice structure described in Ref.~\onlinecite{radzihovsky20vector}.  We expect that the details of how one places $\bA_\beta$ and $\bE_\beta$ on a lattice do not play an important role.

\begin{figure}
        \includegraphics[width=\columnwidth]{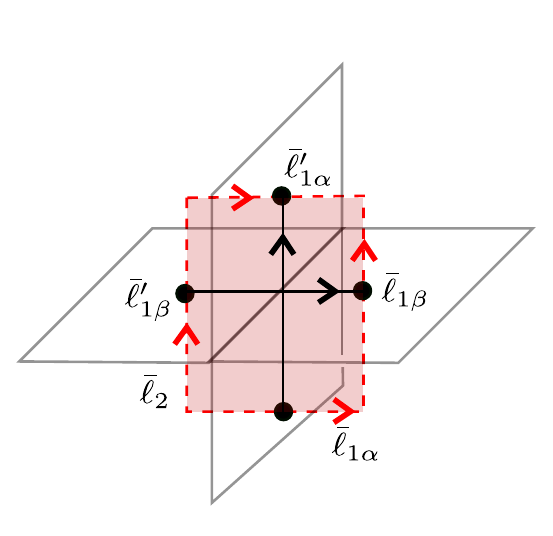}
        \caption{Illustration of the terms involved in $b_{\alpha \beta}(\bar{\ell}_2)$ after gauging (see Eq.~\ref{eqn:bgauged}). The operator $(\bA_\alpha)_\beta$ resides on the vertical link joining $\bar{\ell}_{1 \alpha}$ to $\bar{\ell}'_{1 \alpha}$, with $(\bA_\beta)_\alpha$ on the horizontal link joining $\bar{\ell}'_{1 \beta}$ to $\bar{\ell}_{1 \beta}$.}
        \label{fig:cells2}
\end{figure}

Upon gauging the 0-form symmetries, the magnetic field $b_{\alpha \beta}(\bar{\ell}_2)$ become non-gauge-invariant, and must be modified by
\begin{equation}
\label{eqn:bgauged}
b_{\alpha \beta}(\bar{\ell}_2) \to b_{\alpha \beta}(\bar{\ell}_2) - (\bA_\alpha)_\beta + (\bA_\beta)_\alpha \text{.}
\end{equation}
The notation for the last two terms is somewhat schematic.  In more detail, the penultimate term is $\bA_\alpha$ on a link joining $\bar{\ell}_{1 \alpha}$ to $\bar{\ell}'_{1 \alpha}$, while the last term is
$\bA_\beta$ on a link joining $\bar{\ell}'_{1 \beta}$ to $\bar{\ell}_{1 \beta}$, as illustrated in Fig.~\ref{fig:cells2}. 

The Hamiltonian thus becomes
\begin{eqnarray}
H_{{\rm gauged}}  &=& U \sum_{\ell_{d-1}} [ e_{\alpha}(\ell_{d-1}) ]^2  \nonumber \\
&-& K \sum_{\bar{\ell}_2} \cos [ b_{\alpha \beta} (\bar{\ell}_2 ) - (\bA_\alpha)_\beta + (\bA_\beta)_\alpha   ]  \nonumber \\
&+& U_E \sum \bE_\alpha^2  - K_E \sum \cos (\nabla \times \bA_{\alpha} ) \text{.} \label{eqn:hgauged}
\end{eqnarray}
The last two terms are schematic in form and involve sums over the foliation index $\alpha$ as well as over position.  This Hamiltonian is supplemented by the Gauss law
\begin{equation}
\nabla \cdot \bE_\alpha = e_\alpha(\ell_{d-1}) \text{,} \label{eqn:coupled-u1-gauss}
\end{equation}
where the left-hand side is the lattice divergence of $\bE_\alpha$, which is a sum over the links on which $\bE_\alpha$ is defined that touch $\ell_{d-1}$.

This is precisely the model  constructed and studied on the hypercubic lattice in Ref.~\onlinecite{radzihovsky20vector}, where it was shown that (for $d \geq 3$) when $K$ and $K_E$ are sufficiently large one obtains the scalar-charge theory as a low-energy effective description.  The basic idea is that at energy scales below $K$, the  antisymmetric part of $A_{\alpha \beta} \equiv (\bA_\alpha)_\beta$ is removed, leaving a symmetric-tensor gauge potential.  Another simple observation is that gauge charges of the $e$-field, \emph{i.e.} where $\rho = \nabla \cdot e \neq 0$, are fractons in this theory, since moving such a charge requires creating or removing lines of $e$-field, which themselves carry charge under the $\alpha$ gauge fields.  At the same time, a finite line segment of $e$-field behaves precisely like a dipole in the scalar charge theory, which is free to move, but -- due to the immobility of fractons -- not to stretch, or to rotate and convert into a dipole of different orientation.  

Our construction thus generalizes the scalar charge theory to a general multifoliated manifold and associated cell structure.  However, the scope of generalization that this really allows is not yet clear.  For instance, we do not know examples of non-singular multifoliations with compact leaves on manifolds other than the $d$-torus, or manifolds obtained by cutting and gluing a multifoliated $d$-torus in a manner that respects the foliation structure.  Indeed, Ref.~\onlinecite{slagle19symmetric} has argued that there is an obstruction to defining the scalar charge theory on curved manifolds.  Of course, our construction can be used to study the scalar charge theory on twisted multifoliations of $T^d$ such as that considered in Sec.~\ref{sec:xcube} for the unoriented case, and -- we expect -- on manifolds with singular multifoliations.  

Upon gauging the 0-form symmetries generated by $Q_\beta$, the faithful 1-form symmetry becomes a residual non-faithful 1-form symmetry.  The symmetry operators in the gauged theory are the same as in Eq.~(\ref{eqn:u1-symm-ops}), but now we have $Q_\beta = 0$, where $Q_\beta$ is given by the same expression Eq.~(\ref{eqn:Qbeta}).  This holds simply because the total ${\rm U}(1)$ gauge charge on a closed spatial manifold must vanish.  As noted above, we refer to this non-faithful 1-form symmetry as a framed 1-form symmetry.  The terminology comes from the notion of a framing of a $d$-manifold (also called a parallelization), which is a choice of $d$ smooth vector fields, such that at each point the tangent vectors form a linearly independent set.  Not all manifolds admit a framing, only those whose tangent bundle is isomorphic to a $d$-dimensional trivial vector bundle; a framing can be equivalently understood as a specific choice of trivialization of the tangent bundle.  The reason for this choice of terminology is that, through the $Q_\beta = 0$ constraint, the 1-form symmetry ``knows about'' a choice of $d$ vector fields giving a framing.  This discussion raises the intriguing possibility that a multifoliation structure may not be needed to define the scalar charge theory.  Instead, it may be possible to define the theory on framed spatial manifolds, perhaps even without specifying a lattice structure.  Presumably a Riemannian metric is also needed, in order to define the Maxwell ${\rm U}(1)$ gauge theories used in the construction.

Finally, we discuss breaking of the framed 1-form symmetry. Because the $K$-term creates small loops of $e$-field, these loops are condensed in the large-$K, K_E$ phase, where we expect that the framed 1-form symmetry is broken.  However, at present we have not established this via an analysis like that in Sec.~\ref{sec:breaking}, where in the X-cube model case we showed that certain symmetry operators have non-trivial action on the ground state subspace.  Unlike in that case, here it does not appear to be sufficient to consider $M = T^d$ with suitable generalizations of the twisted foliation structure, as we now explain.  First, suppose $c_f$ is the cellular $(d-1)$-cycle corresponding to a leaf $f$ of the foliation structure.  Then for any two leaves $f, f'$ in the same foliation of $M$, we have $\Phi(c_f) = \Phi(c_{f'})$ within the ground state subspace, because $c_f - c_{f'}$ is a boundary, so that $\Phi(c_f) - \Phi(c_{f'})$  and measures the $e$-charge in the region bounded by the two leaves.    Because $e$-charge excitations are gapped,  this difference vanishes. (More generally, $\Phi(\partial c_d) = 0$ for any $d$-cycle $c_d$, by the same argument.)   Moreover, we also have $\sum_f \Phi(c_f) = Q_\alpha =  0$, where the sum is over all the leaves in the  foliation with index $\alpha$.  It follows that $\Phi(c_f) = 0$ in the ground state subspace.  A consequence is that $\Phi(c_{d-1})$ can only be non-zero within the ground state subspace if $c_{d-1}$ is not homologous to a leaf or sum of leaves.

So far this discussion closely parallels that in the X-cube case, but there is an important difference.  Suppose that $c_{d-1}$ is not homologous to a sum of leaves, but $n c_{d-1}$ is, for some integer $n > 1$.  Then it still must be true that $\Phi(c_{d-1}) = 0$ in the ground-state subspace, simply because $\Phi(n c_{d-1}) = n \Phi(c_{d-1})$.  As an illustration, if one takes $d=3$ and $M = T^3$, choosing an oriented version of the twisted multifoliation, the 2-cycle $c^{xy}_2$ discussed in Sec.~\ref{sec:breaking} and illustrated in Fig.~\ref{fig:symm-op} indeed is not homologous to a sum of leaves, but $2 c^{xy}_2$ is.  While we do not have a rigorous proof, more generally it appears that following the strategy of Sec.~\ref{sec:breaking} will not be successful to diagnose symmetry breaking in this case, if one restricts attention to $M = T^d$ and to obvious generalizations of the twisted multifoliation structure.  This does not mean that the framed 1-form symmetry is unbroken.  It rather means that further study is needed.  For instance, there may exist a multifoliated manifold where some symmetry operator does act non-trivially within the ground state subspace.  Alternatively, it may be possible to study the expectation values of  operators creating charged objects and find a sharp difference in behavior between the small-$K$ and large-$K$ phases, analogous to area-law and perimeter-law behavior in pure gauge theories.

\subsection{$p$-string construction}
\label{sec:u1-pstring}

We now discuss the construction of the scalar charge theory in terms of $p$-string condensation,\cite{radzihovsky20vector} following the outline given in Sec.~\ref{sec:generalized}.  We make use of the same $d$-multifoliation and resulting structure on $M$ discussed above.  While the perspective is different from that in Sec.~\ref{sec:u1-gauging}, the same Hamiltonian arises from the construction, and much of the technical discussion is identical.  Therefore we can be quite brief in this section.

The first step is to introduce $d$ different compact ${\rm U}(1)$ gauge fields, with electric fields $\bE_\alpha$ and vector potentials $\bA_\alpha$, where $\alpha = 1,\dots,d$.  We do not specify the spatial locations where $\bE_\alpha$ and $\bA_\alpha$ reside, but we choose the charge of the $\alpha$ gauge field to reside on $(d-1)$-cells $\ell_{d-1}$ whose foliation index is $\alpha$.  Denoting this charge by $e_\alpha(\ell_{d-1})$, we have the Gauss laws
\begin{equation}
\nabla \cdot \bE_\alpha = e_\alpha(\ell_{d-1}) \text{,}
\end{equation}
just as in Eq.~(\ref{eqn:coupled-u1-gauss}).  The lattice field $e_\alpha(\ell_{d-1})$ takes integer eigenvalues, and we denote the conjugate compact variable by $a_\alpha(\ell_{d-1})$.

We now impose the framed ${\rm U}(1)$ 1-form symmetry exactly as in Eq.~(\ref{eqn:u1-symm-ops}).  We can do this naturally only because of the geometrical structure provided by the multifoliation, which allows us to view $e_\alpha(\ell_{d-1})$ as an integer-valued flux through the $(d-1)$-cell $\ell_{d-1}$.  Without the multifoliation structure, $e_\alpha$ is just a scalar electric charge.

At this point, we need to write down a local, gauge-invariant Hamiltonian that couples together the $d$ ${\rm U}(1)$ gauge theories that we started with, respecting the framed 1-form symmetry.  We are naturally led to $H_{{\rm gauged}}$ of Eq.~(\ref{eqn:hgauged}).  A key point is that the $K$-term of $H_{{\rm gauged}}$ creates and destroys small loops of $p$-string, where the $p$-strings are composed of $e_\alpha$ electric charges.  As shown in Ref.~\onlinecite{radzihovsky20vector} and briefly reviewed in Sec.~\ref{sec:u1-gauging}, the large-$K$ phase is the deconfined phase of the scalar charge theory (for $d \geq 3$).  The present discussion shows that we can interpret this phase as a $p$-string condensate of $e_\alpha$ charge loops.

\section{Discussion}
\label{sec:discussion}

In this paper, we discussed $p$-string condensation, and related constructions involving gauging of symmetry, from the viewpoint of higher form symmetries.  Focusing on the X-cube model and the scalar charge theory, we identified novel types of higher-form symmetry, and showed that the X-cube fracton phase can be understood in terms of breaking a foliated 1-form symmetry.  Here, we briefly comment on some open questions for future work.

Following the discussion in Sec.~\ref{sec:generalized}, we expect it is possible to analyze other coupling and gauging constructions of fracton models in terms of higher-form symmetries.  For instance, Ref.~\onlinecite{radzihovsky20vector} obtained the ${\rm U}(1)$ vector charge theory by coupling together ordinary ${\rm U}(1)$ gauge theories; it should be possible to analyze this construction along the lines of this paper.  While in this paper we focused on 1-form symmetries, 2-form symmetries may also be relevant for fracton phases.  In Ref.~\onlinecite{ma17coupled}, a ``four color cube'' fracton model was obtained by coupling together four X-cube models and condensing membranes built out of the lineon excitations.  Can such $p$-membrane condensation be understood in terms of breaking an exotic 2-form symmetry?

A natural question that arises in this work is the relationship between subsystem symmetry\cite{vijay16fracton} and the exotic higher-form symmetries considered here.  We define a subsystem symmetry to be one with symmetry operators supported on subsystems, which could be \emph{e.g.} planes, lines or even fractals.  This is very general, and higher-form symmetries are clearly a type of subsystem symmetry, but not \emph{vice versa}.  Usually, unlike higher-form symmetries, the subsystem symmetries that play a role in theories of fracton phases involve rigid subsystems, such as lattice planes of a fixed orientation.  Nonetheless there are connections with some of the analysis here; for instance,  in the gauging construction of the X-cube model described in Sec.~\ref{subsec:variants}, the subgroup of the 1-form symmetry that is gauged is a subsystem symmetry, with symmetry operators supported on $xy$, $yz$ and $xz$ planes.  It may be interesting to identify and explore further such connections in future work.

\acknowledgments{We are grateful for discussions and correspondence with Markus Pflaum, Kevin Slagle and Zhenghan Wang.  In addition, MH is particularly grateful for several inspiring discussions with Nathan Seiberg and Shu-Heng Shao.  This research is supported by the U.S. Department of Energy, Office of Science, Basic Energy Sciences (BES) under Award number DE-SC0014415 (MH, and work of MQ prior to September 2019).  The research of MQ is supported by the NDSEG program (starting from September 2019).  The work of LR is supported by a Simons Investigator award from the Simons Foundation.  This work was also partly supported by the Simons Collaboration on Ultra-Quantum Matter, which is a grant from the Simons Foundation (651440, MH).}

\end{document}